\documentclass[%
 aip, 
 amsmath,amssymb,
 reprint,%
]{revtex4-1}

\usepackage{graphicx}
\usepackage{array}
\usepackage{bm}
\usepackage{tabularx}
\usepackage{multirow}
\usepackage[caption=false]{subfig}
\usepackage[utf8]{inputenc}
\usepackage[T1]{fontenc}
\usepackage{mathptmx}
\usepackage{etoolbox}
\usepackage{glossaries}
\makeglossaries
\newacronym{sptt}{sPTT}{simplified Phan-Thien–Tanner}
\newacronym{ml}{ML}{Machine Learning}
\newacronym{nsga2}{NSGA-II}{Non-dominated Sorting Genetic Algorithm II}
\newacronym{ffnn}{FFNN}{Feed-Forward Neural Network}
\newacronym{edl}{EDL}{Electric Double Layer}
\newacronym{pinn}{PINN}{Physics-Informed Neural Network}
\newacronym{svr}{SVR}{Support Vector Regression}
\newacronym{gpr}{GPR}{Gaussian Process Regression}
\newacronym{rf}{RF}{Random Forest Regression}
\newacronym{xgboost}{XgBoost}{eXtreme Gradient Boosting Regression}
\newacronym{mape}{MAPE}{Mean Absolute Percentage Error}
\newacronym{rbf}{RBF}{Radial Basis Function}
\newacronym{nn}{NN}{Neural Network}
\newacronym{rmse}{RMSE}{Root Mean Squared Error}
\newacronym{mxae}{MxAE}{Maximum Absolute Error}

\newcommand{\pb}[1]{\parbox[t]{\dimexpr\textwidth-3.8cm\relax}{\raggedright #1\strut}}

\makeatletter
\def\@email#1#2{%
 \endgroup
 \patchcmd{\titleblock@produce}
  {\frontmatter@RRAPformat}
  {\frontmatter@RRAPformat{\produce@RRAP{*#1\href{mailto:#2}{#2}}}\frontmatter@RRAPformat}
  {}{}
}%
\makeatother
\begin{document}

\preprint{AIP/123-QED}

\title[]{Data-Driven Design Optimization of Streaming-Potential-Mediated Electrokinetic Transport of Viscoelastic Fluids in Microchannels}
% Force line breaks with \\
\author{Ankan Basu}
\affiliation{%
  Department of Computer Science and Engineering, Jadavpur University, Kolkata, India 700032%\\This line break forced% with \\
  }%
 \altaffiliation[Also at ]{India Meteorological Department, Kolkata, India}%Lines break automatically or can be forced with \\
\author{Sumanta Banerjee}%
 \email{sumanta.banerjee@heritageit.edu}
\affiliation{ 
Department of Mechanical Engineering, Heritage Institute of Technology, Kolkata, India 700107%\\This line break forced with \textbackslash\textbackslash
}%

\author{Sandip Sarkar}
\affiliation{%
Department of Mechanical Engineering, Jadavpur University, Kolkata, India 700032%\\This line break forced% with \\
}%

\date{\today}% It is always \today, today,
             %  but any date may be explicitly specified

\begin{abstract}
Streaming-potential-mediated transport of viscoelastic fluids has attracted research attention owing to its applications in electrokinetic energy conversion and microfluidic transport. Existing analytical and semi-analytical models in published literature provide valuable physical insights, but require repeated numerical evaluations for exploring large design spaces and identifying the optimal operating conditions. In this work, a surrogate-assisted framework is developed for rapid design optimization of pressure-driven electrokinetic transport of simplified Phan-Thien-Tanner fluids in a slit microchannel. A high-fidelity numerical database is generated over a broad range of governing dimensionless parameters, which includes the zeta potential, the Debye parameter, the Dukhin number, and the viscoelastic parameter. A Neural Network surrogate model is subsequently trained to accurately approximate the nonlinear relationship between the governing parameters and the streaming potential, while the volumetric flow rate and hydroelectric energy conversion efficiency were calculated from closed form equation by using the streaming potential predicted by the surrogate. This is coupled with a multi-objective optimization strategy to identify operating conditions that simultaneously maximize energy conversion efficiency and volumetric flow rate. The proposed methodology can significantly accelerate parametric exploration compared with repeated numerical simulations across different parameters and provides practical design guidelines for electrokinetic microfluidic devices. The study demonstrates the potential of combining computational fluid mechanics with data-driven surrogate modeling for efficient engineering design and optimization.
\end{abstract}


\maketitle

\section{Introduction}

Micro- and nanofluidic devices can convert part of the electrochemical energy stored at a charged solid-liquid interface into usable electrical work whenever a pressure-driven flow displaces the mobile ions of the \gls{edl}. This streaming-potential-mediated conversion mechanism has been proposed as a route to powering lab-on-a-chip diagnostics, implantable biosensors, and other low-power devices operating on physiological or otherwise complex fluids. Such fluids---including biological fluids (e.g., blood, saliva, DNA solutions) and synthetic polymer-bearing biofluids---routinely display viscoelastic rather than Newtonian rheology. A realistic account of streaming-potential transport in these systems must couple electrokinetics with a constitutive model capable of capturing shear-thinning and normal-stress effects. To isolate this fundamental coupling, we employ the simplified Phan-Thien-Tanner (sPTT) model [1]. We adopt the standard assumptions for dilute polymer solutions in narrow confinements: the solvent viscosity is assumed negligible relative to the polymeric contribution (\(\eta_{\text{eff}} = 0\)), and the fluid properties are treated as temperature-independent. These simplifications allow us to focus exclusively on the polymeric stress dynamics and neglect the effects of solvent relaxation or thermal gradients. Furthermore, the governing dimensionless parameters are restricted to physically relevant ranges: \(\zeta \in [1, 10]\), \(\kappa \in [10, 40]\), \(\epsilon_s W_{ik}^2 \in [0,2]\), and \(Du \in [0,25]\). At physiological ionic strengths (where the Debye length is \(\sim 1\) nm), these \(\kappa\) values correspond to nanochannel half-heights on the order of \(10-40\) nm. While physiological zeta potentials are typically a few tens of millivolts, the optimal design regime identified later (\(\zeta \approx 6\), or \(\sim 150\) mV) is readily realizable in highly charged synthetic nanofluidic channels (e.g., silica or track-etched nanopores) or by operating at very dilute ionic strengths (\(<0.1\) mM) where microchannel half-heights of \(\sim 0.1-1\) \(\mu\)m align with \(\kappa \in [10, 40]\).

Building on this constitutive foundation, closed-form and semi-analytical solutions for electrokinetically driven \gls{sptt} flow have been derived under a range of forcing conditions: electro-osmotic flow under the Debye-Hückel approximation~\cite{afonso2009}, combined electro-osmotic/pressure-driven flow with several viscoelastic constitutive models~\cite{ferras2016}, electro-osmotic flow at high zeta potentials beyond the Debye-Hückel regime~\cite{sarma2018}, and non-isothermal electro-osmotic transport with Joule heating and Soret effects~\cite{hernandez2023}, and, for other viscoelastic and microstructured fluids, time-periodic electro-osmotic flows of Jeffreys~\cite{eosf2026} and Brinkman--micropolar~\cite{elsapa2025} fluids. 

Most directly relevant to the present work, Sarkar~\cite{sarkarSpttStreamingPot} solved the full nonlinear Poisson-Boltzmann equation, without invoking the Debye-Hückel linearization, to obtain exact expressions for the induced streaming potential 
$E$, the velocity field, the volumetric flow rate $q$, the polymeric stress components, and the hydroelectric energy conversion efficiency $\eta$ of a pressure-driven \gls{sptt} flow, all parametrized by four governing dimensionless groups: the wall zeta potential $\zeta$, the ratio of channel half-height to Debye length $\kappa$, the Dukhin number $Du$, and a compound viscoelastic parameter $\epsilon_s W_{i_k}^2$.

None of these studies, however, perform a systematic search over the full four-dimensional parameter space for the operating point -- or the trade-off surface of operating points — that maximizes device performance. Reference~\cite{sarkarSpttStreamingPot} identifies, by inspection of two-parameter contour plots with the remaining variables held fixed, that $q$ and $\eta$ both appear to peak near $\zeta \approx 6$ and grow with $\epsilon_s W_{i_k}^2$; this is useful qualitative guidance, but it falls short of a global optimum, since  $\kappa$ and $Du$ are never varied jointly with $\zeta$ and $\epsilon_s W_{i_k}^2$, and flow rate and efficiency are never formally posed as competing objectives. Closing this gap directly is complicated by the structure of the governing equations themselves: $E$ is defined only implicitly, through an zero-net-current condition that must be solved iteratively, and the associated velocity and flow-rate expressions require evaluating polylogarithm series whose convergence demands fine quadrature resolution precisely in the thin-\gls{edl}, large-$\kappa$ regime where the physics is most consequential. Embedding this pipeline directly inside a population-based multi-objective optimizer, which typically requires many thousands of objective-function evaluations, is computationally expensive.

In this work, we remove that bottleneck with a \gls{ml} surrogate. We generate a dataset of $( \zeta, Du, \kappa, \epsilon_s W_{i_k}^2) \xrightarrow{} E$ pairs by iteratively solving the zero-net-current condition of Ref.~\cite{sarkarSpttStreamingPot} to convergence at each sampled point, and train a \gls{ffnn} to reproduce this mapping. Because $q$ and $\eta$ remain available in closed form once $E$ is known, the surrogate replaces only the expensive implicit step. We then couple the trained surrogate with the \gls{nsga2}~\cite{deb2002nsga2} to jointly maximize $q$ and $\eta$ across the full physically relevant range of all four dimensionless groups simultaneously, producing a Pareto-optimal front of designs rather than a single, locally inferred estimate. The remainder of this paper is organized as follows: the next section reviews the relevant electrokinetic–viscoelastic and \gls{ml} literature and situates the present contribution; the governing equations are then presented; this is followed by a description of the surrogate architecture, training procedure, and the \gls{nsga2} optimization framework. Thereafter, the results are presented discussing the variation of streaming potential, volumetric flow rate, and efficiency with different input parameters along with resulting Pareto front and its design implications. The final section reports the conclusions and future scope of work.

It is essential to note the scope of the optimal conditions identified in this study (\(\kappa \approx 10\), \(\zeta \approx 6\), \(Du \approx 0\), \(\epsilon_s W_{ik}^2 \approx 2\)). While the model assumes zero solvent viscosity, real biofluids possess a finite solvent fraction. As discussed in the limitations of the underlying sPTT model (and the base formulation in Ref. [8]), a finite solvent viscosity acts to dampen the shear-thinning augmentation of the velocity profile and shifts the effective viscosity ratio. Consequently, the absolute magnitudes of the flow rate and efficiency reported here represent an upper bound; however, the qualitative trade-off structure and the Pareto-optimal trends are expected to persist, albeit with reduced sensitivity to the viscoelastic parameter. The optimal design space identified here therefore serves as a fundamental baseline for dilute viscoelastic electrolytes in highly confined nanochannels. The integration of finite solvent viscosity and specific real-fluid parameter mappings are discussed further in the Limitations section and deferred to future work.

\section{Literature Review}

\subsection{Electrokinetic transport of viscoelastic fluids}
The theoretical study of electrokinetically driven non-Newtonian flow has matured considerably since Afonso, Alves, and Pinho~\cite{afonso2009} obtained analytical solutions for electro-osmotic flow of FENE-P and linear PTT fluids under the Debye–Hückel approximation. Ferrás et al.~\cite{ferras2016} extended this framework to closed-form and semi-analytical solutions spanning several viscoelastic constitutive models under combined electro-osmotic and pressure-driven forcing, and Sarma et al.~\cite{sarma2018} later derived exact solutions for PTT electro-osmotic flow valid at high zeta potentials, moving beyond the Debye-Hückel restriction that limited earlier treatments. The general net zero current formalism used to close the streaming-potential problem — and the associated hydroelectric conversion-efficiency metric — was itself established for Newtonian electrolyte flow by Munshi and Chakraborty~\cite{munshi2009} and for oscillatory Maxwell-fluid forcing by Bandopadhyay and Chakraborty~\cite{bandopadhyay2012}, before Sarkar~\cite{sarkarSpttStreamingPot} extended it to pressure-driven \gls{sptt} flow under the full Poisson–Boltzmann equation. That model has continued to generate follow-on work from the same group, including an extension to thermally developing flow with Joule heating and viscous dissipation~\cite{kumbalpuri2024} and, most recently, a hybrid skip-connected \gls{pinn} for the same base geometry~\cite{basu2025}.

\subsection{Machine learning for electrokinetic and viscoelastic flow modeling}
Neural surrogates and \glspl{pinn} have begun to appear alongside, rather than in place of, this semi-analytical machinery. Reference~\cite{basu2025} applies a \gls{pinn} directly to the pressure driven and electro-osmotic flow of \gls{sptt} fluid system used in the present study, but for heat transfer analysis instead of streaming potential. In the related but distinct context of the Poisson-Boltzmann equation itself, Achondo, Chaudhry, and Cooper~\cite{achondo2025} recently benchmarked \gls{pinn} architectures for molecular electrostatics and found that careful input/output scaling and Fourier-feature embeddings were necessary to reach competitive accuracy. This serves as a reminder that even the electrostatic half of this class of problem is nontrivial to learn directly.

\subsection{Multi-objective design optimization}
Coupling a fast neural-network surrogate to a population-based multi-objective optimizer, most often \gls{nsga2}~\cite{deb2002nsga2}, is a well-established recipe for engineering design problems in which multiple objectives must be traded off across a moderate-dimensional parameter space, and it has been applied productively to thermally analogous microchannel transport problems. This combination has not, to our knowledge, been applied to the electrokinetic-viscoelastic streaming-potential problem, where flow rate and hydroelectric conversion efficiency are natural design objectives.

\subsection{Gaps and present contribution}

Three gaps follow from the above. First, existing analytical treatments of streaming-potential-mediated \gls{sptt} flow, including Ref.~\cite{sarkarSpttStreamingPot} itself, explore the four-dimensional parameter space only through fixed-partial sweeps -- varying one or two of $(\zeta, Du, \kappa, \epsilon_s W_{i_k}^2)$ while holding the remaining variables at a single representative value. So, the reported ``optimal" regions are qualitative and locally inferred rather than the outcome of a joint global search. Second, the zero-net-current condition fixing $E$ is implicit and requires iterative solution together with polylogarithm-heavy quadrature that becomes numerically demanding at large $\kappa$, making direct embedding of the semi-analytical model inside a many-evaluation optimizer such as \gls{nsga2} expensive. Third, although both $q$ and $\eta_s$ have independently been proposed as design targets in this literature, they have not been formally posed as competing objectives so as to map the trade-off -- or lack of one -- between them.

The present work addresses all three gaps. We 
\begin{enumerate}
    \item generate training data for $(\zeta, Du, \kappa, \epsilon_s W_{i_k}^2)$ by iteratively solving the zero-net-current condition of Ref.~\cite{sarkarSpttStreamingPot} across the physically relevant ranges established in that study ($\zeta \in [1, 10]$, $\kappa \in [10, 40]$, $\epsilon_s W_{i_k}^2 \in [0, 2]$, $Du \in [0, 25]$), and train a \gls{ffnn} surrogate for this mapping;

    \item couple the trained surrogate, together with the closed-form expressions for $q$ and $\eta_s$ that follow once $E$ is known, to \gls{nsga2}, enabling a genuinely joint search over all four dimensionless groups simultaneously; and

    \item report the resulting Pareto-optimal design set for flow rate and conversion efficiency across the full design space in place of the single, fixed-parameter optimum region identified qualitatively in prior work.
\end{enumerate}

\section{Methods}

\subsection{Governing Equations}

\begin{figure}[htbp]
    \centering
    \includegraphics[width=0.9\columnwidth]{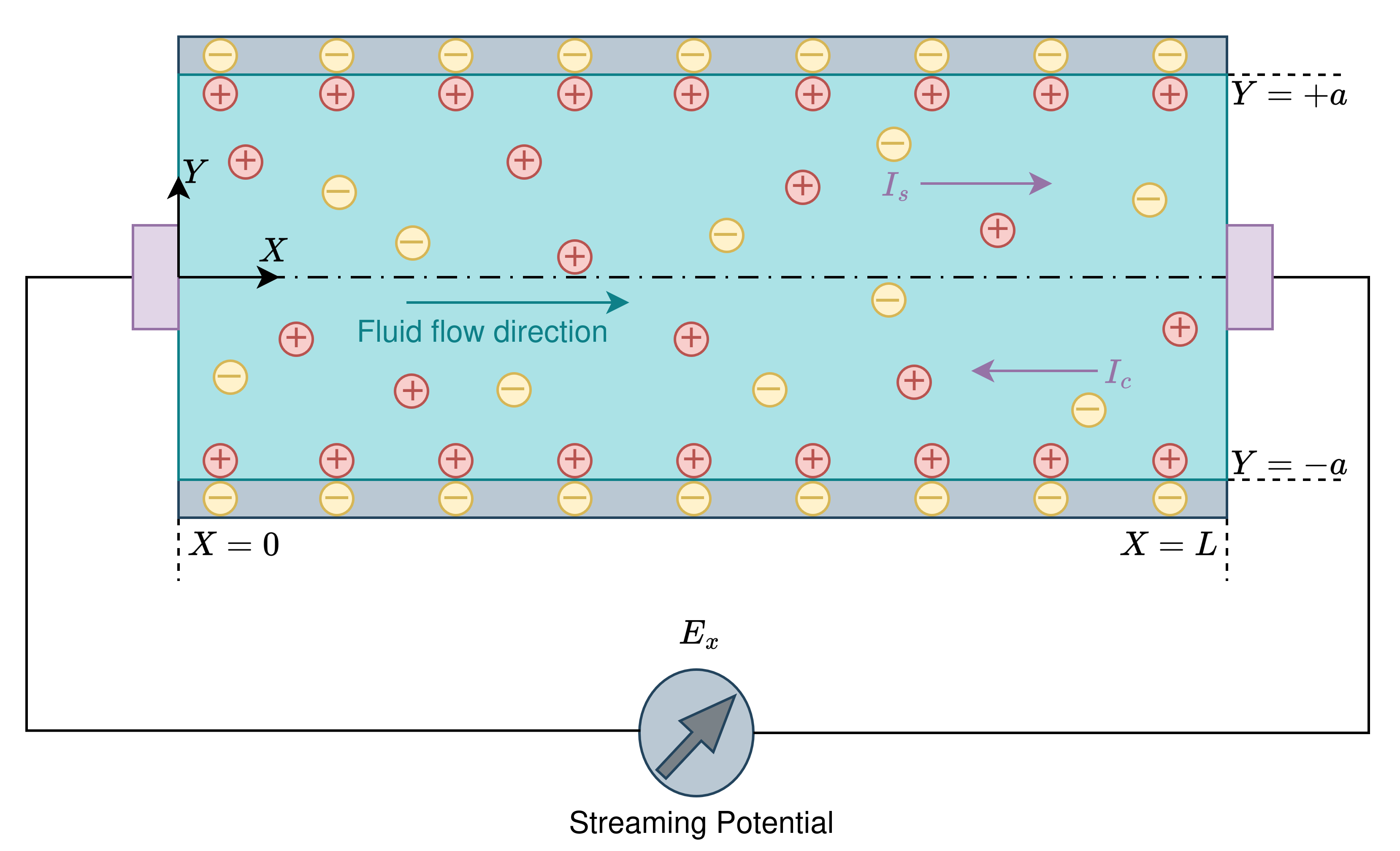}
    \caption{Schematic diagram of the flow geometry under consideration}
    \label{fig:flow_config}
\end{figure}

This work considers the electrokinetic flow of a non-Newtonian ionic fluid (modeled as the \gls{sptt} fluid) through a slit-type parallel plate microchannel. The width of the microchannel is $W$, length is $L$ and separation between the plates is $2a$, where $W \ll L$. The origin is chosen to be located at the centreline of the channel with the X-axis running horizontally and the Y-axis running vertically, such that the two plates of the channel are located at Y-coordinate $\pm a$, the beginning of the channel is at X-coordinate $0$ and the channel end is at X-coordinate $L$. This configuration is illustrated in Fig.~\ref{fig:flow_config}.

The channel walls acquire a uniform surface potential $\zeta^*$ through the formation of an \gls{edl}, and the electrolyte is taken to be $z:z$ symmetric with permittivity $\epsilon$. No electric field is imposed externally, but the motion of the fluid drags the mobile counter-ions of the \gls{edl} downstream, and the resulting charge separation sets up a back-directed field $E_X$ -- the streaming potential -- through the competition between advective transport and electromigration. The two \glspl{edl}, at $Y=+a$ and $Y=-a$, are taken to be thin and non-overlapping, so each develops independently of the other; the flow itself is steady, unidirectional, and incompressible, with fluid properties held constant and independent of temperature throughout.

\subsubsection{Electrostatic Potential Distribution}

The electrostatic potential $\Psi$ induced within each \gls{edl} satisfies Poisson's equation, (Eq.~\eqref{eqn:poisson}), where $\rho_e$ is the local net charge density. For the fully developed, unidirectional flow considered here, with thin and non-overlapping \glspl{edl}, this reduces to the one-dimensional form (Eq.~\eqref{eqn:poisson-steady}).

\begin{equation} \label{eqn:poisson}
    \nabla^2 \Psi = - \frac{\rho_e}{\epsilon}
\end{equation}

\begin{equation} \label{eqn:poisson-steady}
    \frac{d^2 \Psi}{d Y^2} = - \frac{\rho_e}{\epsilon}
\end{equation}

Neglecting steric effects, the local charge density follows a Boltzmann distribution, (Eq.~\ref{eqn:boltzman-dist}) where symbol $e$ is elementary charge, $k_B$ is Boltzmann's constant, $T$ is the absolute temperature, $n_0$ is the bulk ionic concentration, and $z$ is the ionic valence.

\begin{equation} \label{eqn:boltzman-dist}
    \rho_e = - 2 n_0 e z \sinh \left( \frac{e z \Psi}{k_B T} \right)
\end{equation}

Combining Eq.~\eqref{eqn:poisson-steady} and Eq.~\eqref{eqn:boltzman-dist} gives the governing Poisson--Boltzmann equation, Eq.~\eqref{eqn:poisson-boltzman}, where $1/\bar\kappa^2 = \epsilon k_BT/(2n_0e^2z^2)$ is the square of the Debye length.

\begin{equation} \label{eqn:poisson-boltzman}
    \frac{d^2 \Psi}{dY^2} = \bar{\kappa}^2 \sinh \left( \frac{e z \Psi}{k_B T} \right)
\end{equation}

Rather than linearizing Eq.~\eqref{eqn:poisson-boltzman} via the Debye--H\"uckel approximation, which valid only for low surface potentials, its full nonlinear form is retained and integrated directly, subject to the centreline symmetry condition $d\Psi/dY|_{Y=0}=0$ and the wall condition $\Psi|_{Y=a}=\zeta^*$, giving the closed-form, dimensionless \gls{edl} potential distribution -- Eq.~\eqref{eqn:edl-pot-dimensionless}, where $\psi=ez\Psi/k_BT$, $y=Y/a$, $\zeta=ez\zeta^*/k_BT$, and $\kappa=a\bar\kappa$.

\begin{equation} \label{eqn:edl-pot-dimensionless}
    \psi = 4 \tanh^{-1} \left( \tanh \left( \frac{\zeta}{4} \right) e^{- \kappa (1 - y)} \right)
\end{equation}

It is this full-scale solution, rather than its Debye--H\"uckel-linearized counterpart, that is carried through the remainder of the analysis and used later to generate the training data for the neural-network surrogate.

\subsubsection{Governing Flow Equations and the sPTT Constitutive Model}

The fluid motion is governed by the continuity and Cauchy momentum equations, Eq.~\eqref{eqn:continuity} and Eq.~\eqref{eqn:cauchy-momentum}, where $\tau$ is the polymeric extra-stress tensor and $\eta_{eff}$ is the Newtonian solvent viscosity, taken to be negligible relative to the polymeric contribution ($\eta_{eff}=0$) throughout this work.

\begin{equation} \label{eqn:continuity}
    \nabla \cdot \vec{U} = 0
\end{equation}

\begin{equation} \label{eqn:cauchy-momentum}
    \rho \frac{D \vec{U}}{D t} = - \nabla p + \nabla \cdot \tau + \eta_{eff} \nabla^2 \vec{U} + \vec{F}_E
\end{equation}

The source term $\vec F_E$ is the electrokinetic body force arising from the induced streaming-potential field,

\begin{equation} \label{eqn:streaming-pot-field}
    \vec{F}_{E_X} = \rho_e E_X
\end{equation}

which couples the electrostatics of the preceding subsection back into the momentum balance.

The polymeric stress is closed with the \gls{sptt} constitutive model \cite{thien1977}, written compactly as Eq.~\eqref{eqn:stress-tensor}, where $f(\mathrm{tr}\,\tau)$ is a scalar stress-coefficient function, $\lambda$ is the fluid relaxation time, $\eta_p$ is the polymeric viscosity, $D$ is the rate-of-deformation tensor, and $\bar{\bar\tau}$ is the Gordon--Schowalter convected derivative of the stress (Eq.~\eqref{eqn:gordon-schowalter}).

\begin{equation} \label{eqn:stress-tensor}
    f(tr(\tau)) \tau + \lambda \bar{\bar{\tau}} = 2 \eta_p D
\end{equation}

\begin{equation} \label{eqn:gordon-schowalter}
    \bar{\bar{\tau}} = \frac{D \tau}{D t} - \left( \nabla \vec{U}^T \cdot \tau + \tau \cdot \nabla \vec{U} \right)
\end{equation}

For the sPTT model, $f$ takes the affine linear form

\begin{equation} \label{eqn:gordon-schowalter-sptt}
    f(tr(\tau)) = 1 + \frac{\epsilon_s \lambda}{\eta_p} \tau_{kk},
\end{equation}

with $\epsilon_s$ the extensibility parameter bounding the extensional viscosity and $\tau_{kk}$ the trace of the stress tensor. Specializing to fully developed, unidirectional flow, $\vec U \equiv \{U(Y),0,0\}$, with two-dimensional stress ($\tau_{ZZ}=0$), the constitutive relation reduces to

\begin{equation} \label{eqn:stress-tensor-2}
    f(\tau_{XX} + \tau_{YY}) \tau_{XX} = 2 \lambda \frac{dU}{dY} \tau_{XY},
\end{equation} 

\begin{equation} \label{eqn:stress-tensor-3}
    f(\tau_{XX} + \tau_{YY}) \tau_{YY} = 0,
\end{equation}

\begin{equation} \label{eqn:shear-stress}
    f(\tau_{XX} + \tau_{YY}) \tau_{XY} = (\eta_p + \lambda \tau_{YY}) \frac{dU}{dY}.
\end{equation}

Equation \eqref{eqn:stress-tensor-3} gives $\tau_{YY}=0$ directly, so $\tau_{kk}=\tau_{XX}$; dividing Eq.~\eqref{eqn:stress-tensor-2} by Eq.~\eqref{eqn:shear-stress} then gives $\tau_{XX}=(2\lambda/\eta_p)\tau_{XY}^2$, and substituting this through Eq.~\eqref{eqn:gordon-schowalter-sptt} back into Eq.~\eqref{eqn:shear-stress} collapses the system to a single cubic relation between shear stress and velocity gradient:

\begin{equation} \label{eqn:stress-tensor-4}
    \tau_{XY} + \frac{2 \epsilon_s \lambda^2 (\tau_{XY})^3}{\eta_p^2} = \eta_p \frac{dU}{dY}
\end{equation}

which reduces to Newton's law of viscosity in the limit $\epsilon_s \to 0$.

\subsubsection{Velocity Field}

Under the same steady, fully developed, incompressible assumptions, the streamwise momentum equation reduces to

\begin{equation} \label{eqn:cauchy-momentum-2}
    \frac{d \tau_{XY}}{d Y} = \frac{dp}{dX} + \epsilon E_X \frac{d^2 \Psi}{d Y^2}
\end{equation}

where the electrokinetic body force has been written, via Poisson's equation, in terms of the curvature of the EDL potential. Applying centreline symmetry, $[\tau_{XY},\Psi]_{Y=0}=0$, and integrating once gives

\begin{equation} \label{eqn:cauchy-mometum-integrated}
    \tau_{XY} = \frac{dp}{dX} Y + \epsilon E_X \frac{d \Psi}{dY}.
\end{equation}

Substituting this into the constitutive relation, Eq.~\eqref{eqn:stress-tensor-4}, gives an explicit, nonlinear equation for the velocity gradient,

\begin{equation} \label{eqn:flow-velocity-pde}
    \frac{dU}{dY} = \frac{1}{\eta_p} \left[ \frac{dp}{dX} Y + \epsilon E_X \frac{d \Psi}{dY} \right] + \frac{2 \epsilon_s \lambda^2}{\eta_p^3} \left[ \frac{dp}{dX} Y + \epsilon E_X \frac{d \Psi}{d Y} \right]^3,
\end{equation}

which is nondimensionalized using $u=U/U_{ref}$, $U_{ref}=-(a^2/\eta_p)(dp/dX)$, $E=E_X/E_{ref}$, $E_{ref}=-(a^2/\epsilon\zeta^*)(dp/dX)$, and the Weissenberg number $Wi_k=\lambda\kappa U_{ref}/a$, giving the governing dimensionless ODE

\begin{equation} \label{eqn:flow-velocity-pde-nondim}
    \frac{du}{dy} = \left( \frac{E}{\zeta} \frac{d \psi}{dy} - y \right) + \frac{2 \epsilon_s {W_{i_k}}^2}{\kappa^2} \left( \frac{E}{\zeta} \frac{d \psi}{dy} - y \right)^3.
\end{equation}

Equation \eqref{eqn:flow-velocity-pde-nondim} admits a closed-form solution under the no-slip conditions $u(y)|_{y=\pm1}=0$ and centreline symmetry, given in full by Eq.~\eqref{eqn:velocity-integrated}, where $p=\tanh(\zeta/4)$ and $\mathrm{Li}_s(\cdot)$ denotes the polylogarithm (Jonqui\`ere) function of order $s$ \cite{sarkarSpttStreamingPot}.

\begin{widetext}
\begin{equation} \label{eqn:velocity-integrated}
\begin{aligned}
    u(y) &= \frac{24 \epsilon W_{i_k^2} E}{\kappa^4 \zeta} \left[ Li_3(p e^{\kappa (y-1)}) - Li_3(-p e^{\kappa (y-1)}) + Li_3(-p) - Li_3(p) \right]  \\
    &\quad + \frac{24 \epsilon W_{i_k^2} E}{\kappa^3 \zeta} \left[ y Li_2(-p e^{\kappa (y-1)}) -y Li_2(p e^{\kappa (y-1)}) + Li_2(p) - Li_2(-p) \right] \\ 
    &\quad - \frac{24 \epsilon W_{i_k^2} E}{\kappa^2 \zeta^2} \left[ \ln\left( \frac{1 - p^2 e^{2 \kappa (y -1 )}}{1 - p^2} \right) - \frac{y^2 \zeta}{2E} \ln\left( \frac{1 + pe^{\kappa (y - 1}}{1 - pe^{\kappa (y - 1}} \right) - \frac{\zeta}{2E} \ln\left( \frac{1 - p}{1 + p} \right) \right] \\ 
    &\quad + \frac{16 \epsilon_s W_{i_k^2} E^3 p}{\zeta^3} \left[ \frac{2 e^{\kappa (y + 3)}}{\left( e^{2 \kappa} - p^2e^{2 \kappa y} \right)^2} - \frac{e^{\kappa (y + 1)}}{e^{2 \kappa} - p^2e^{2 \kappa y}} -\frac{\left( 1 + p^2 \right)}{\left( 1 - p^2 \right)^2} \right] \\
    &\quad - \frac{48 \epsilon_s W_{i_k^2} E^2}{\kappa \zeta^2} \left[ \frac{y e^{2 \kappa}}{e^{2 \kappa} - p^2e^{2 \kappa y}} - \frac{1}{1 - p^2} + (1-y) \right] \\ 
    &\quad - \frac{16 \epsilon_s W_{i_k^2} E^3}{\zeta^3} \left[ \tanh^{-1} (p e^{\kappa (y-1)}) - \tanh^{-1}(p)  \right] \\
    &\quad + \frac{\epsilon_s W_{i_k^2}}{2 \kappa^2} \left( 1 - y^4 \right) \\ 
    &\quad + \frac{4E}{\zeta} \left[ \tanh^{-1} (p e^{\kappa (y-1)}) - \tanh^{-1}(p)  \right] \\ 
    &\quad + \frac{1}{2} (1 - y^2)
\end{aligned}
\end{equation}
\end{widetext}

This closed form is the quantity evaluated at every iteration of the net zero current solve described next, and is therefore the principal computational cost that motivates the surrogate model.

\subsubsection{Zero-Net-Current and the Streaming Potential}

The streaming potential $E$ appearing throughout Eqs.~\eqref{eqn:flow-velocity-pde-nondim}--\eqref{eqn:velocity-integrated} is not known in advance: it is fixed only implicitly, by the requirement that the streaming current generated by advection of the \gls{edl}'s mobile charge be exactly balanced by the conduction current returning through the bulk electrolyte and the Stern layer. Dimensionally,

\begin{equation} \label{eqn:electro-neutrality}
    \int_{-a}^a z e (n_{+} - n_{-}) U(Y)dY + \frac{z^2 e^2 E_x}{f} \int_{-a}^a (n_{+} + n_{-}) dY + 2 \sigma_{stern} E_X = 0,
\end{equation}

where the first term is the streaming current, the second is the bulk conduction current (with $f=2n_0e^2z^2/\sigma_B$ the ionic friction factor), and the third is the conduction current through the Stern layer, of conductivity $\sigma_{stern}$. In dimensionless form, with the Dukhin number $Du=\sigma_{stern}/(a\sigma_B)$ and conductivity parameter $\alpha=\eta\sigma_B/(2\epsilon n_0 k_BT)$, this becomes

\begin{equation} \label{eqn:electro-neutrality-nondim}
    \int_0^1 u(y) \sinh (\psi) dy - \frac{E \alpha}{\zeta} \left( \int_0^1 \cosh (\psi) dy + Du \right) = 0.
\end{equation}

Because $u(y)$ in the first integral is itself a function of $E$ through Eq.~\eqref{eqn:velocity-integrated}, Eq.~\eqref{eqn:electro-neutrality-nondim} is nonlinear in $E$ and must be solved iteratively -- guess $E$, evaluate $u(y)$, update $E$ from Eq.~\eqref{eqn:electro-neutrality-nondim}, and repeat to convergence -- separately at every point in the four-dimensional design space $(\zeta,\kappa,\epsilon_sWi_k^2,Du)$. It is this repeated per-point iteration, compounded by the cost of evaluating the polylogarithm series in Eq.~\eqref{eqn:velocity-integrated} at high $\kappa$, that the neural-network surrogate is trained to replace.

\subsubsection{Flow Rate and Energy-Conversion Efficiency}

Once $E$ is known, the dimensionless volumetric flow rate follows by integrating the velocity profile across the channel:

\begin{equation} \label{eqn:flow-rate}
    Q = W \int_{-a}^a U(Y) dY,
\end{equation}

\begin{equation} \label{eqn:flow-rate-nondim}
    q = \frac{Q}{2aWU_{ref}} = \int_0^1 u(y) dy.
\end{equation}

This quantity, $q$, is the first of the two objectives maximized in the multi-objective optimization described in the next section.

The second objective is the hydroelectric energy-conversion efficiency, the ratio of electrical power delivered by the streaming current and streaming potential to the mechanical power supplied by the applied pressure gradient:

\begin{equation} \label{eqn:efficiency}
    \eta = \frac{I_s |E_s|}{|\frac{dp}{dX}|Q}.
\end{equation}

Nondimensionalizing this ratio using the same reference scales introduced above gives

\begin{equation} \label{eqn:efficiency-nondim}
    \eta = \left( \frac{\kappa^2}{\zeta} \right) \frac{i_s |E|}{q}
\end{equation}

where $i_s$ is the dimensionless streaming current obtained from the first term of Eq.~\eqref{eqn:electro-neutrality-nondim}. Equations \eqref{eqn:flow-rate-nondim} and \eqref{eqn:efficiency-nondim} together define the two objectives, $q$ and $\eta$, over which the multi-objective search of the following section is performed.

\subsection{Surrogate Architecture} \label{sec:surrogate-arch}

Multiple surrogate models were developed --- \gls{rf}, \gls{xgboost}, \gls{svr}, \gls{gpr}, and a \gls{nn}. Their respective characteristics and configurations are summarized in Table~\ref{tab:surrogates}.

\begin{table*}[ht]
    \centering
    \caption{Surrogate models and their configurations.}
    \label{tab:surrogates}
    \renewcommand{\arraystretch}{1.3}
    \begin{ruledtabular}
    \begin{tabular}{ll}
        \textbf{Surrogate} & \textbf{Model architecture and hyperparameters} \\
        \hline

        RF & \pb{500 decision trees, unrestricted tree depth, minimum samples per split = 2, minimum samples per leaf = 1, bootstrap sampling enabled.} \\

        XGBoost & \pb{500 decision trees, with a maximum tree depth of 4 and a learning rate of 0.05.} \\

        SVR &
        \pb{Two kernel configurations --- linear and RBF. Both used a regularization parameter of $C=1.0$ and an $\varepsilon$-insensitive loss parameter of $\varepsilon=0.1$. The parameter $C$ controls the trade-off between model complexity and training error, while $\varepsilon$ determines the width of the error-insensitive tube within which prediction errors are not penalized.} \\

        GPR &
        \pb{Two covariance kernel configurations --- RBF and Matérn ($\nu=1.5$). Both configurations incorporated a White kernel to model observation noise. Kernel hyperparameters were optimized by maximizing the log marginal likelihood.} \\

        Neural Network &
        \pb{A fully connected neural network with three hidden layers and GELU activation functions. The input layer comprises four neurons corresponding to the four input parameters. The output layer consists of a single neuron with a Softplus activation function. The network architecture is illustrated in Fig.~\ref{fig:nn_arch}. The Adam optimizer was used for training, with a \texttt{ReduceLROnPlateau} learning-rate scheduler. Model was trained till the error plateaued on the training set.} \\

        \end{tabular}
    \end{ruledtabular}
\end{table*}

The surrogate models take $\kappa$, $\epsilon_s W_{i_k^2}$, $\zeta$, and $Du$ as input and output the predicted nondimensional streaming potential $E$. The inputs were scaled using z-score normalization with the mean and standard deviation calculated using the training set and subsequently applied to the test set. This preprocessing was performed for all models except \gls{rf} and \gls{xgboost}, as tree-based models are generally insensitive to differences in feature scales and parameter ranges making z-score normalization unnecessary.

\begin{figure}[htbp]
    \centering
    \includegraphics[width=0.9\columnwidth]{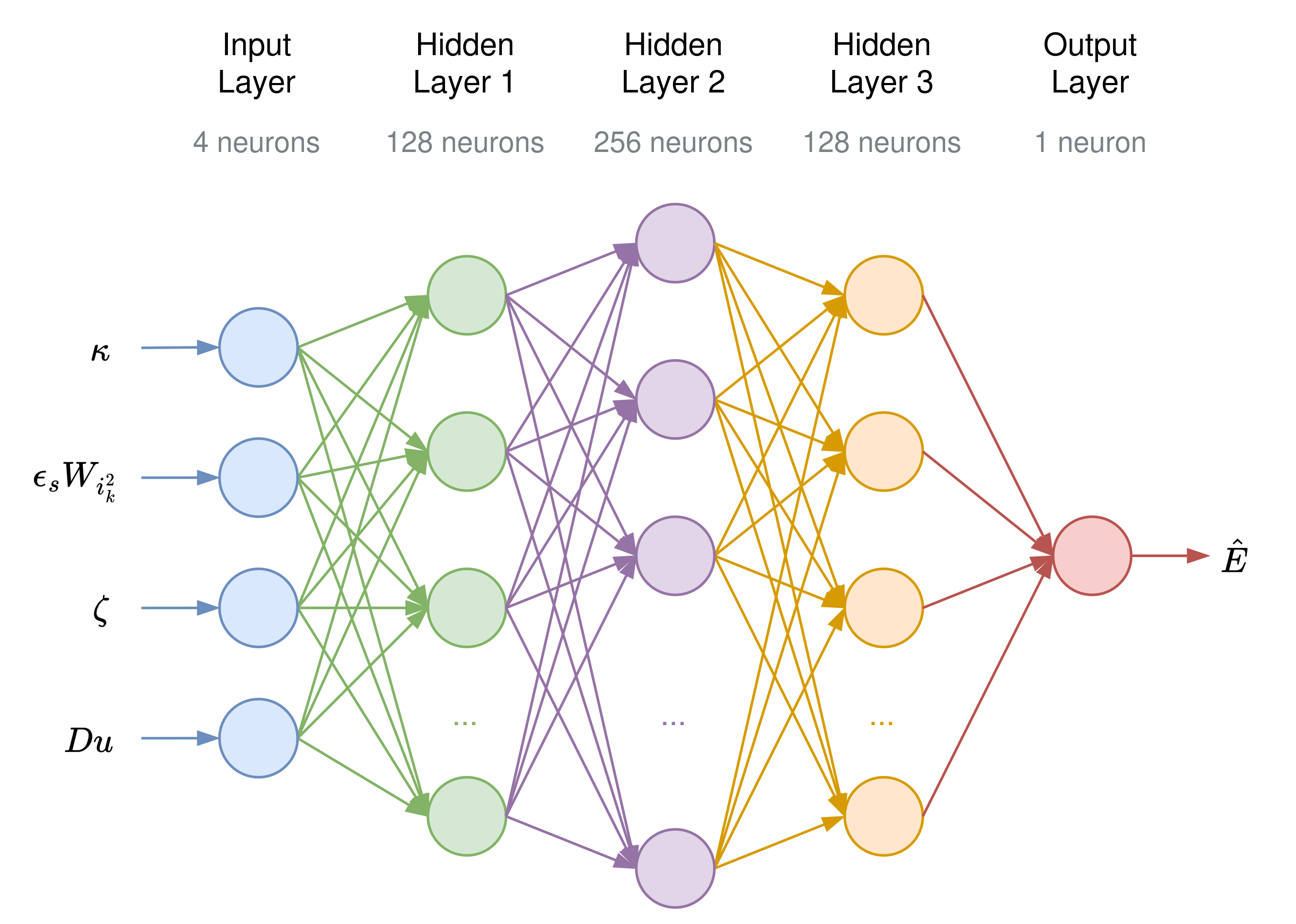}
    \caption{Architecture of the surrogate neural network}
    \label{fig:nn_arch}
\end{figure}

Eqs.~\eqref{eqn:velocity-integrated} and~\eqref{eqn:electro-neutrality-nondim} were solved to generate the data for training the network. 5000 combinations of the parameters $\kappa$, $\epsilon_s W_{i_k^2}$, $\zeta$, and $Du$ were generated in the range $[10, 40]$, $[0, 2]$, $[1, 10]$, and $[0, 25]$ respectively by Latin Hypercube Sampling. Values of nondimensional streaming potential $E$ were calculated for the chosen combinations of $\kappa$, $\epsilon_s W_{i_k^2}$, $\zeta$, and $Du$. The value of $\alpha$ in Eq.~\eqref{eqn:electro-neutrality-nondim} was taken as 10 (keeping in line with the work of Sarkar~\cite{sarkarSpttStreamingPot}). The training dataset, thus, contained 5000 input--output data points. Another 1000 data points were generated with a fresh run of Latin Hypercube Sampling with a different seed. This was used as the holdout/test dataset.

The L1 loss computed on log-scaled data was used as the training loss function. Since the nondimensional streaming potential spans a wide range of values, a logarithmic transformation was applied to the target variable to reduce its dynamic range and facilitate model training. If $\hat{E}_i$ is the predicted streaming potential for the $i$-th data point in the dataset, $E_i$ is the corresponding actual streaming potential, and $N$ is the total number of data points, then the loss function can be defined as:

\begin{equation} \label{eqn:log-l1}
    L1 \ Loss =\frac{1}{N} \sum_{i=1}^{N} |\log(\hat{E}_i) - \log(E_i)|
\end{equation}

For model evaluation, L1 Error and \gls{mape} was used. No logarithmic transformation was applied when calculating these evaluation metrics, ensuring that the errors were expressed in terms of the original physical quantity. The evaluation metrics can be defined as:

\begin{equation} \label{eqn:l1-error}
    L1 \ Error =\frac{1}N{} \sum_{i=1}^{N} |\hat{E}_i - E_i|
\end{equation}

\begin{equation} \label{eqn:percentage-error}
    MAPE = \frac{1}{N} \sum_{i=1}^{N} \frac{|\hat{E}_i - E_i|}{\max(|E_i|, \delta)} \times 100
\end{equation}

Where the symbols have same meaning as in Eq.~\ref{eqn:log-l1}. In Eq.~\eqref{eqn:percentage-error}, $\delta$ is a small positive constant used to prevent division by zero. During computation it was set to the smallest positive normal value representable in the \texttt{float64} data type $\delta \approx 2.22507\times10^{-308}$.

\subsection{Optimization Problem Formulation}

The design problem is posed as the simultaneous maximization of two performance metrics -- the dimensionless volumetric flow rate $q$ and the hydroelectric energy conversion efficiency $\eta$ over the four-dimensional design vector $(\zeta, \kappa, \epsilon_s W_{i_k}^2, Du)$, bounded by the physically relevant ranges ($\zeta \in [1, 10]$, $\kappa \in [10, 40]$, $\epsilon_s W_{i_k}^2 \in [0, 2]$, $Du \in [0, 25]$) as identified in Ref.~\cite{sarkarSpttStreamingPot}:

\begin{equation}
    maximize \ F(x) = [q(x), \eta(x)], \ x \in [x_{min}, x_{max}] \subset \mathbb{R}^4
\end{equation}

Because $q$ and $\eta$ need not be maximized by the same $x$, no single point necessarily dominates the entire design space. The goal is therefore to recover the Pareto-optimal set -- designs for which no objective can be improved without degrading the other.

\gls{nsga2}~\cite{deb2002nsga2} is a population-based evolutionary algorithm developed to solve problems of this kind without requiring the objectives to be combined into a single weighted scalar in advance. In this work, each candidate design is encoded as a chromosome of the four decision variables; an initial population of $200$ candidates is evaluated and evolved over $100$ generations through three mechanisms:

\begin{enumerate}
    \item \textbf{Fast non-dominated sorting:} The population is ranked into successive Pareto ``fronts". Designs not dominated by any other design constitute \texttt{Front 1}, designs dominated only by \texttt{Front 1} members form \texttt{Front 2}, and so on. A design $x_a$ is said to dominate $x_b$ if it is at least as good in both objectives and strictly better in at least one. For example, if $q(x_a) \geq q(x_b)$ and $\eta(x_a) \geq \eta(x_b)$, with at least one strict inequality, then the design $x_a$ is dominant over $x_b$.

    \item \textbf{Crowding-distance estimation:} Within each front, a crowding distance is computed for every member from the spacing of its neighbors in objective space. This measure preserves diversity along the front by favoring designs in sparsely populated regions.

    \item \textbf{Elitist selection:} Parent and offspring populations are pooled, and the next generation is filled by taking successive fronts, highest-ranked first, until the population size is reached; if a front must be split partway through, crowding distance breaks the tie. Offspring are generated from selected parents via simulated binary crossover (probability 0.9, distribution index $\eta_c = 15$) and polynomial mutation (probability 0.9, distribution index $\eta_m = 20$)~\cite{pymoo}.
\end{enumerate}

\gls{nsga2} is evaluation-hungry by construction: a population of $N$ individuals evolved over $G$ generations requires on the order of $N \times G$ objective-function ($F(x)$) evaluations. Evaluating $F(x)$ directly at each of these points would require iterating the zero-net-current condition to convergence and evaluating the associated polylogarithm series at sufficient quadrature resolution -- a cost that grows sharply at large $\kappa$. Instead, $F(x)$ is evaluated by first querying a trained surrogate ($\hat{E}(x)$) for the streaming potential, then substituting this prediction into the closed-form expressions for $q$ and $\eta$. This preserves the analytical structure of the model everywhere except the single implicit step the surrogate replaces, and reduces the cost of each objective-function evaluation from an iterative solve to a single forward pass through the network. The search bounds $[x_{min}, x_{max}]$ were kept within the parameter range discussed earlier ($\zeta \in [1, 10]$, $\kappa \in [10, 40]$, $\epsilon_s W_{i_k}^2 \in [0, 2]$, $Du \in [0, 25]$).

\section{Results and Discussion}

\subsection{Surrogate Results} \label{sec:res-surrogate}

The surrogate models were trained and tested on the same training and testing sets and evaluated using the protocols described in Section~\ref{sec:surrogate-arch}. The error metrics and training time are reported in Table~\ref{tab:surrogate-res}.

\begin{table*}[htbp]
\caption{Error values of the surrogate neural network on the training and validation datasets}
\label{tab:surrogate-res}
\begin{ruledtabular}
\begin{tabular}{ lccccr } 
 \multirow{2}{*}{Model} & \multicolumn{2}{c}{Training} & \multicolumn{2}{c}{Testing} & \multirow{2}{*}{Training Time} \\ 
 & L1 Error & MAPE & L1 Error & MAPE & \\
    \hline
 RF & 1.51978e-05 & 2.08450 & 3.82368e-05 & 5.24078 & 137.38549 secs \\
 XgBoost & 2.18687e-05 & 3.49847 & 2.97645e-05 & 4.82704 & 0.63483 secs  \\
 SVR - Linear & 2.59540e-04 & 43.74868 & 2.96012e-04 & 43.93064 & 1.79196 secs  \\
 SVR - RBF & 3.82971e-05 & 5.72728 & 4.13608e-05 & 6.04974 & 0.34636 secs  \\
 GPR - Matérn & 3.26022e-10 & 0.00005 & 2.84463e-06 & 0.38979 & 1284.02303 secs \\
 GPR - RBF & 5.11278e-10 & 0.00011 & 1.74000e-06 & 0.1837 & 2491.97892 secs \\
 NN & 1.61099e-06 & 0.22694 & 1.75379e-06 & 0.24422 & 182.23869 secs \\

\end{tabular}
\end{ruledtabular}
\end{table*}

From Table~\ref{tab:surrogate-res} it can be seen that the \gls{nn} surrogate provides the best balance between error and training time. The \gls{rbf}-\gls{svr}, \gls{rf}, and \gls{xgboost} models exhibit comparatively higher prediction errors, with Linear-\gls{svr} yielding the poorest performance. This suggests that the underlying relationship between $(\zeta, \kappa, \epsilon_s W_{i_k}^2, Du)$ and $E$ to be highly non-linear. The \gls{gpr} models performed very strongly on the training set but showed significant overfitting when their test set error is calculated. Furthermore, they require the longest training times among the evaluated models, which is consistent with the $O(N^3)$ computational complexity of conventional exact \gls{gpr} training, where $N$ denotes the number of training samples (here, 5000). The \gls{nn} surrogate showed strong performance --- with the test set loss very close to that of the \gls{gpr} models while exhibiting no significant deterioration between training and test performance and requiring substantially less training time. Considering these factors, the \gls{nn} model was selected as the surrogate for the subsequent analyses presented in this manuscript.

To assess the computational advantage of the surrogate model, the evaluation times of the NN surrogate and the original numerical solver were compared over 5000 individual evaluations. The findings are summarized in Table~\ref{tab:eval-time}.

\begin{table}[htbp]
\centering
\caption{Computational evaluation time of the NN surrogate and the numerical solver for 5000 individual evaluations.}
\label{tab:eval-time}
\begin{ruledtabular}
\begin{tabular}{lc}
Model & Total Evaluation Time (sec) \\
\hline
NN Surrogate & 1.28422 \\
Numerical Solver & 165.86981 \\
\end{tabular}
\end{ruledtabular}
\end{table}

For a fair comparison, both methods were evaluated sequentially, with one input sample processed at a time (repeated for 5000 times). Although the \gls{nn}-surrogate supports batch inference, this capability was deliberately not utilized, for a fair comparison, because the numerical solver was not coded to support batch evaluation. Even under this sequential evaluation setting, the \gls{nn}-surrogate required only 1.28422 s, compared with 165.86981 s for the numerical solver, corresponding to a speedup of approximately 129 times. This substantial reduction in evaluation time highlights the computational advantage of the surrogate model. Furthermore, since batch inference was not employed, the reported speedup represents a conservative estimate of the potential computational benefit of the \gls{nn}-surrogate when evaluating large numbers of candidate designs during optimization (during which normal batch inference was used).

All simulations, surrogate model training, and multi-objective optimization experiments were performed using the Kaggle cloud computing environment. The computational environment specifications are summarized in Table~\ref{tab:system-spec}. 

\begin{table}[htbp]
\centering
\caption{Computational environment and software specifications}
\label{tab:system-spec}
\begin{ruledtabular}
\begin{tabular}{lr}
Specification & Details \\
\hline
OS & Ubuntu 22.04.5 LTS (Linux Kernel 6.18.55+) \\
CPU &  Intel Xeon CPU @ 2.00 GHz \\
RAM & 32GB \\
GPU & Nvidia Tesla T4 (15GB VRAM) \\ 
\end{tabular}
\end{ruledtabular}
\end{table}

The selected \gls{nn} surrogate model outputs the dimensionless streaming potential ($E$) for dimensionless input parameters $\kappa$, $\epsilon_s W_{i_k^2}$, $\zeta$, and $Du$. From this predicted streaming potential ($E$), the dimensionless volumetric flow rate ($q$) and efficiency ($\eta$) were calculated using Eq.~\eqref{eqn:flow-rate-nondim} and Eq.~\eqref{eqn:efficiency-nondim} respectively. 

Since the surrogate directly predicts only $E$, errors in the predicted streaming potential may also affect the derived quantities --- $q$ and $\eta$. To assess this effect, both quantities were calculated using the streaming potential predicted by the \gls{nn} surrogate (for data points in the holdout/test dataset) and the corresponding value obtained from the numerical solver. The same governing expressions were used in both calculations (Eq.~\eqref{eqn:flow-rate-nondim} and Eq.~\eqref{eqn:efficiency-nondim}). The resulting errors in $E$, $q$, and $\eta$, quantified using the L1 error (Eq.~\eqref{eqn:l1-error}) and \gls{mape} (Eq.~\eqref{eqn:percentage-error}), are reported in Table~\ref{tab:surrogate-err-prop}.

\begin{table*}[htbp]
\caption{Prediction errors in the streaming potential and the corresponding errors propagated to the derived volumetric flow rate and efficiency as calculated for input data points from the test set}
\label{tab:surrogate-err-prop}
\begin{ruledtabular}
\begin{tabular}{ lcccccc } 
 \multirow{2}{*}{Comparison} & \multicolumn{2}{c}{$E$} & \multicolumn{2}{c}{$q$} & \multicolumn{2}{c}{$\eta$} \\ 
 & L1 Error & \% Error & L1 Error & \% Error & L1 Error & \% Error \\
\hline

 Mean & 1.75379e-06 & 0.24422 & 1.71620e-06 & 0.00051 & 4.50609e-06 & 0.18574 \\
 Median & 2.96969e-07 & 0.11342 & 2.91601e-07 & 0.00008 & 8.46760e-07 & 0.10980 \\
 90th percentile & 2.95915e-06 & 0.38418 & 2.90100e-06 & 0.00088 & 7.53879e-06 & 0.37333 \\
 99th percentile & 2.90482e-05 & 1.41740 & 2.77940e-05 & 0.00840 & 7.87584e-05 & 1.41715 \\
\end{tabular}
\end{ruledtabular}
\end{table*}

From Table~\ref{tab:surrogate-err-prop} it can be seen that the streaming potential is predicted with a mean absolute error of $1.75379\times10^{-6}$. Although $E$ spans more than three orders of magnitude over the design space (as seen in Fig.~\ref{fig:E_vs_Du}, Fig.~\ref{fig:E_vs_W}, and Fig.~\ref{fig:E_vs_zeta}), the relative error is small throughout: the median pointwise error is $0.11342\%$, the mean $0.24422\%$, the 90th percentile $0.38418\%$ and the 99th percentile $1.41740\%$. Thus, from the Table~\ref{tab:surrogate-err-prop}, it can be seen that approximately 99\% of the test samples exhibit relative errors below $1.41\%$ and absolute error below $2.90482\times10^{-6}$, indicating consistently accurate predictions across the test set.

The mean absolute percentage error in the volumetric flow rate is only $0.00051\%$, with the 99th percentile remaining below $0.009\%$. This indicates that the prediction errors in $E$ have a negligible effect on the calculated flow rate over the test set. The corresponding \gls{mape} in efficiency is $0.18574\%$, with a median of $0.10980\%$ and a 99th percentile of $1.41715\%$. Thus, although the propagation of prediction errors is more pronounced for $\eta$ than for $q$, the resulting errors remain small across the majority of the test samples. This, along with the results from Table~\ref{tab:eval-time}, demonstrates the suitability of using the surrogate neural network for the subsequent multiobjective optimization and sensitivity analyses

The variation of $E$, $q$, and $\eta$ with the different input parameters are discussed in the following subsections.

\subsubsection{Variation of Streaming Potential with the Dukhin Number}

\begin{figure}[htbp]
    \centering
    \includegraphics[width=0.9\columnwidth]{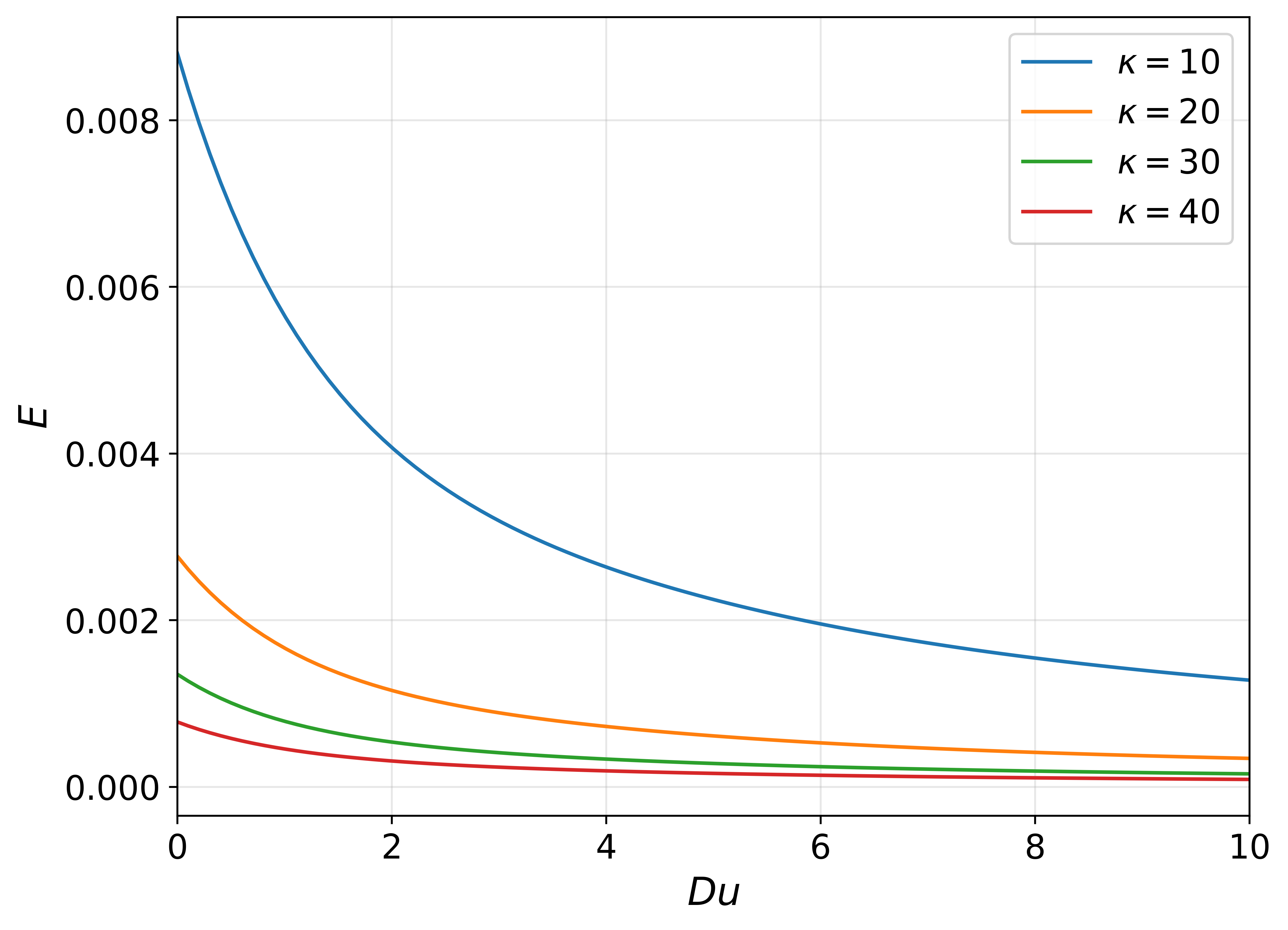}
    \caption{$E$ vs $Du$ plot from the surrogate outputs for $\epsilon_s Wi_k^2 = 1.5$, $\zeta = 4$, and $\kappa = 10, 20, 30, 40$.}
    \label{fig:E_vs_Du}
\end{figure}

Figure~\ref{fig:E_vs_Du} illustrates the dependence of the induced, non-dimensional  streaming potential \(E\) on the Stern layer conductivity, as quantified by the Dukhin number \(Du = \sigma_{\text{Stern}}/(a\sigma_B)\), across different values of the Debye parameter \(\kappa\) (the latter represents the ratio of channel half-height to Debye length). For all values of \(\kappa\), Fig.~\ref{fig:E_vs_Du} shows that \(E\) exhibits a monotonic exponential decay with increasing values of \(Du\). Crucially, this decay is hyperbolic rather than exponential, as \(E \propto 1/(C+Du)\), where \(C \approx 1 + \frac{4}{\kappa}\sinh^2(\zeta/4)\) is the bulk conduction integral.

This behavior can be understood by examining the zero-net-current condition (refer to Eq.~\eqref{eqn:electro-neutrality-nondim}). In the equation, the Dukhin number \(Du\) appears explicitly in the denominator of the conduction current term. As \(Du\) increases, the immobile layer of ions adsorbed directly to the wall (Stern layer) provides an increasingly conductive parallel pathway for the conduction current. This reduces the electrical resistance of the system, implying that a lower magnitude of streaming potential is sufficient to balance the streaming current and satisfy net zero current. The Stern layer essentially acts as a ``short circuit" that bypasses the bulk electrolyte and diminishes the induced electric field. Higher Stern layer conductivity enables more efficient charge relaxation through surface conduction, suppressing the development of the streaming potential. This trend persists across all \(\kappa\) values, although the absolute magnitude of \(E\) decreases with increasing \(\kappa\). This is because thinner \glspl{edl} (for larger \(\kappa\)) contain fewer mobile ions to generate the streaming current.

\subsubsection{Variation of Volumetric Flow Rate with the Dukhin Number}

\begin{figure}[htbp]
    \centering
    \includegraphics[width=0.9\columnwidth]{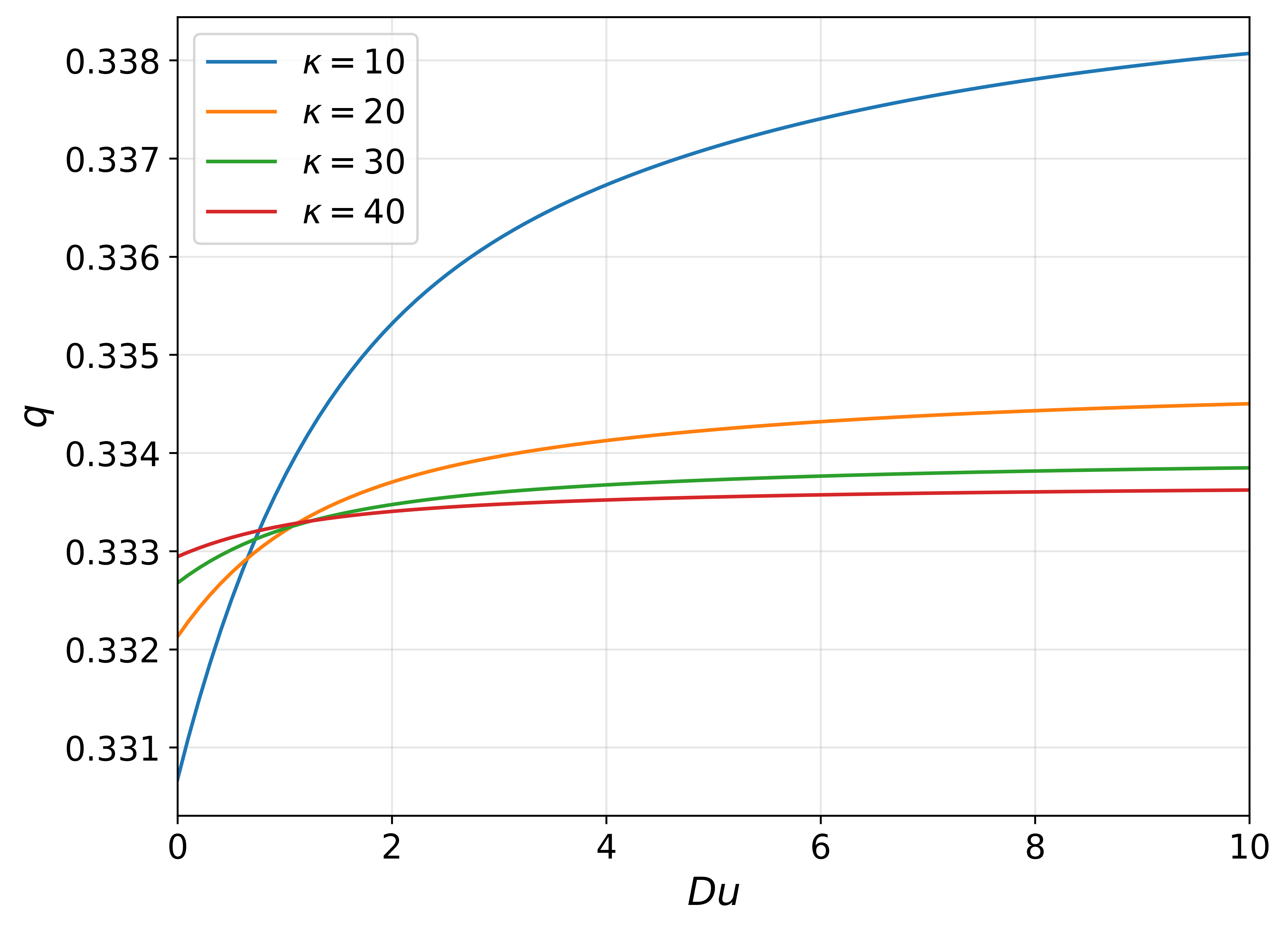}
    \caption{$q$ vs $Du$ plot from the surrogate outputs for $\epsilon_s Wi_k^2 = 1.5$, $\zeta = 4$, and $\kappa = 10, 20, 30, 40$.}
    \label{fig:q_vs_Du}
\end{figure}

Figure~\ref{fig:q_vs_Du} demonstrates that the flow rate \(q\) increases monotonically with the Dukhin number \(Du\) for all values of \(\kappa\), approaching the zero-field value \(q_0\) from below. The curves for different \(\kappa\) cross near \(Du \approx 1\) at \(q \approx 1/3\).

The Dukhin number represents the ratio of Stern layer conductivity to bulk conductivity. As \(Du\) increases, the highly conductive Stern layer provides a short-circuit path, suppressing the streaming potential \(E \propto 1/(C+Du)\). Since \(E\) generates an electrokinetic body force that opposes the pressure-driven flow, a reduction in \(E\) diminishes this retarding force, which allows \(q\) to rise. At \(Du=0\), the retardation costs \(2.6\%\) of \(q_0\) for \(\kappa=10\), whereas at larger \(Du\), \(q\) approaches \(q_0\). The crossing near \(Du \approx 1\) arises from \textit{two competing effects} that both scale as \(1/\kappa^2\): the \textit{shear-thinning gain} (\(q_0 - 1/3 = 2\epsilon_s Wi_k^2/(5\kappa^2)\)) and the \textit{retardation loss} (\(q_0 - q \approx E\)). At small \(Du\), the loss dominates, making the thick-EDL channel (\(\kappa=10\)) having the lowest flow rate. At larger \(Du\), the retardation is short-circuited, and the larger shear-thinning gain of the \(\kappa=10\) channel makes it the highest.

\subsubsection{Variation of Energy Conversion Efficiency with the Dukhin Number}

\begin{figure}[htbp]
    \centering
    \includegraphics[width=0.9\columnwidth]{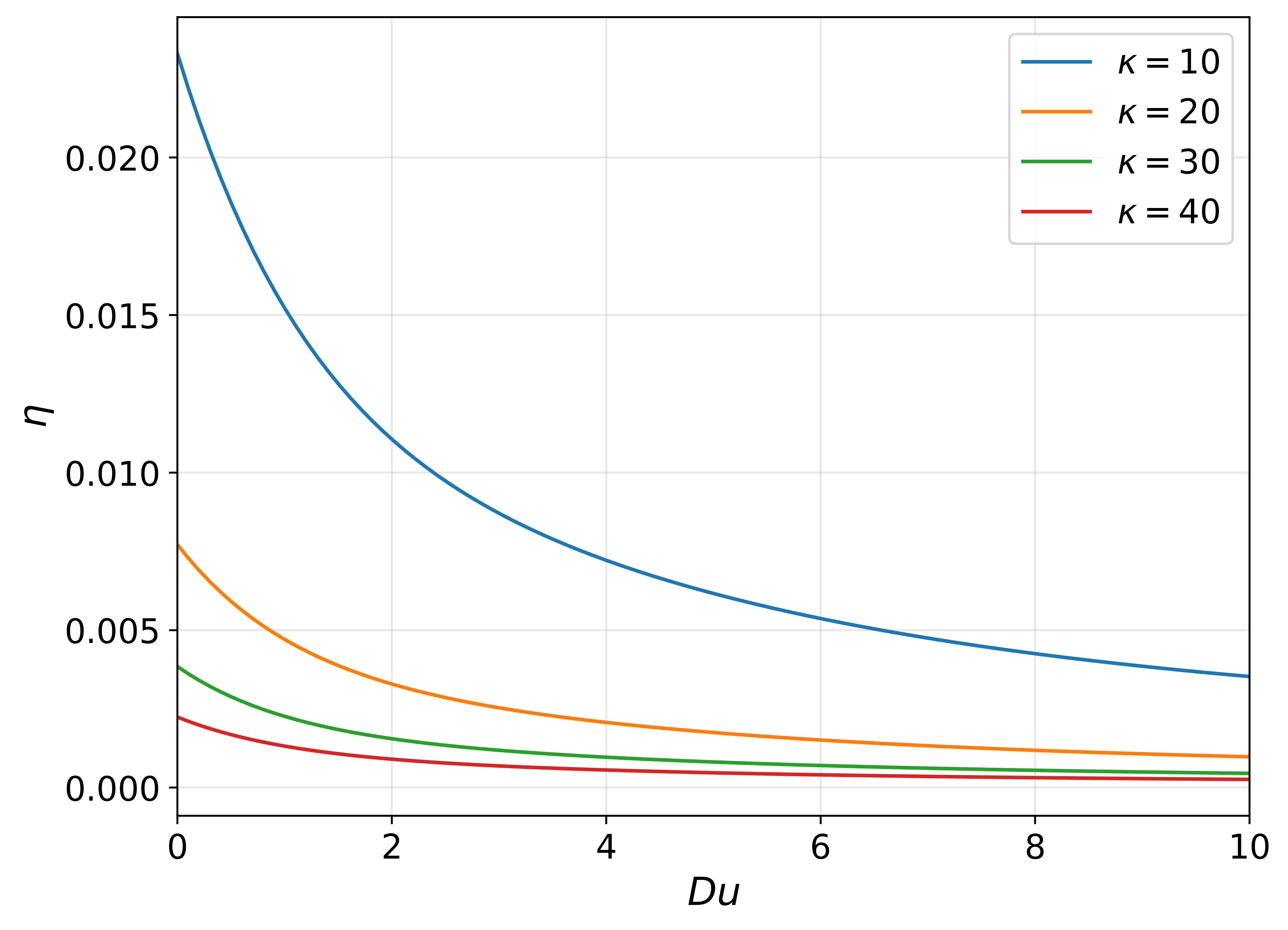}
    \caption{$\eta$ vs $Du$ plot from the surrogate outputs for $\epsilon_s Wi_k^2 = 1.5$, $\zeta = 4$, and $\kappa = 10, 20, 30, 40$.}
    \label{fig:eta_vs_Du}
\end{figure}

Figure~\ref{fig:eta_vs_Du} shows that the efficiency \(\eta\) decreases monotonically with increasing \(Du\), displaying a hyperbolic decay that closely mirrors the behavior of \(E\).

The conversion efficiency is directly tied to the magnitude of the streaming potential. As the \(Du\) increases, the Stern layer short-circuits the system and drastically reduces \(E\). Over this sweep, the streaming current \(i_s\) changes by less than about \(7\%\) and \(q\) by about \(2\%\). This relates \(\eta \propto E\) to a good approximation. As the Stern layer carries current but does no electrical work, it lowers the electrical power \(i_s E\) while leaving the hydraulic power nearly unchanged. Thus, efficiency is highest when the Stern-layer conduction is negligible (\(Du=0\)) and falls by an order of magnitude from \(\kappa=10\) to \(\kappa=40\), since both \(i_s\) and \(E\) scale as \(1/\kappa^2\).

\subsubsection{Variation of Streaming Potential with Viscoelastic Parameter}

\begin{figure}[htbp]
    \centering
    \includegraphics[width=0.9\columnwidth]{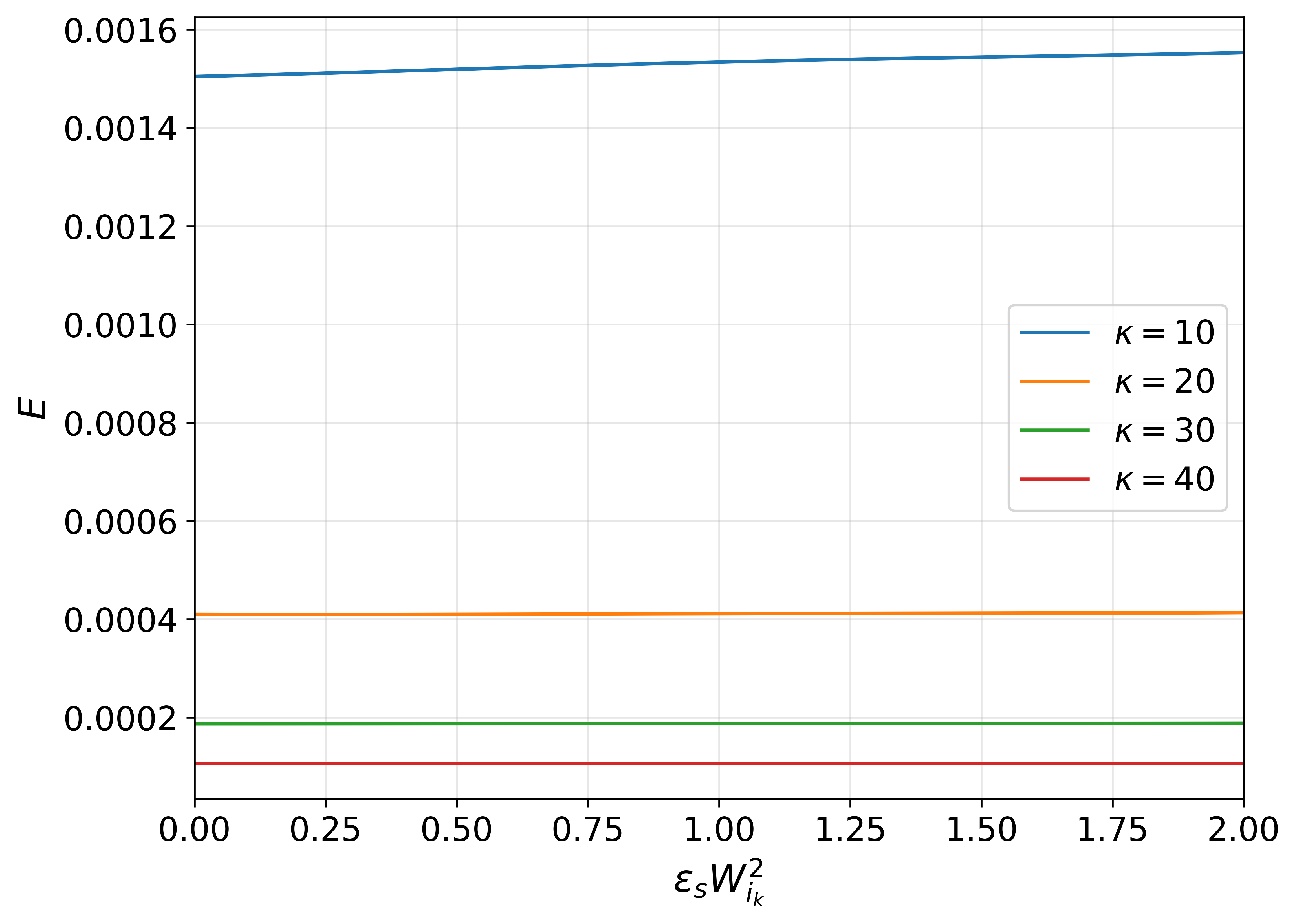}
    \caption{\(E\) vs \(\epsilon_s Wi_k^2\) plot from the surrogate outputs for \(\zeta = 4\), \(Du = 8\), and \(\kappa = 10, 20, 30, 40\)}
    \label{fig:E_vs_W}
\end{figure}

Figure~\ref{fig:E_vs_W} shows that the streaming potential \(E\) is essentially \textit{independent} of the viscoelastic parameter \(\epsilon_s Wi_k^2\) across all values of \(\kappa\), exhibiting only a very weak variation (about \(4-5\%\) for \(\kappa=10\), and less for larger \(\kappa\)).

The zero-net-current condition dictates that \(E = (\zeta/\alpha) i_s / (C+Du)\). Increasing the value of \(\epsilon_s W_{i_k}^2\) significantly enhances shear-thinning, and the corresponding correction to the velocity profile is \(\Delta u = (c/4)(1-y^4)\), with \(c = 2\epsilon_s Wi_k^2/\kappa^2\). This correction vanishes at the wall (\(y=\pm 1\)) and is largest at the centerline. However, the streaming current integral \(i_s = \int_0^1 u \sinh\psi \, dy\) is weighted by \(\sinh\psi\), which is concentrated in the thin \gls{edl} near the walls. Because the extra velocity from shear-thinning accumulates in the core rather than the \gls{edl}, \(i_s\) changes by at most a few percent. As the conduction integral \(C\) does not depend on \(\epsilon_s W_{i_k}^2\) at all, the driving force for \(E\) remains nearly constant.

\subsubsection{Variation of Volumetric Flow Rate with Viscoelastic Parameter}

\begin{figure}[htbp]
    \centering
    \includegraphics[width=0.9\columnwidth]{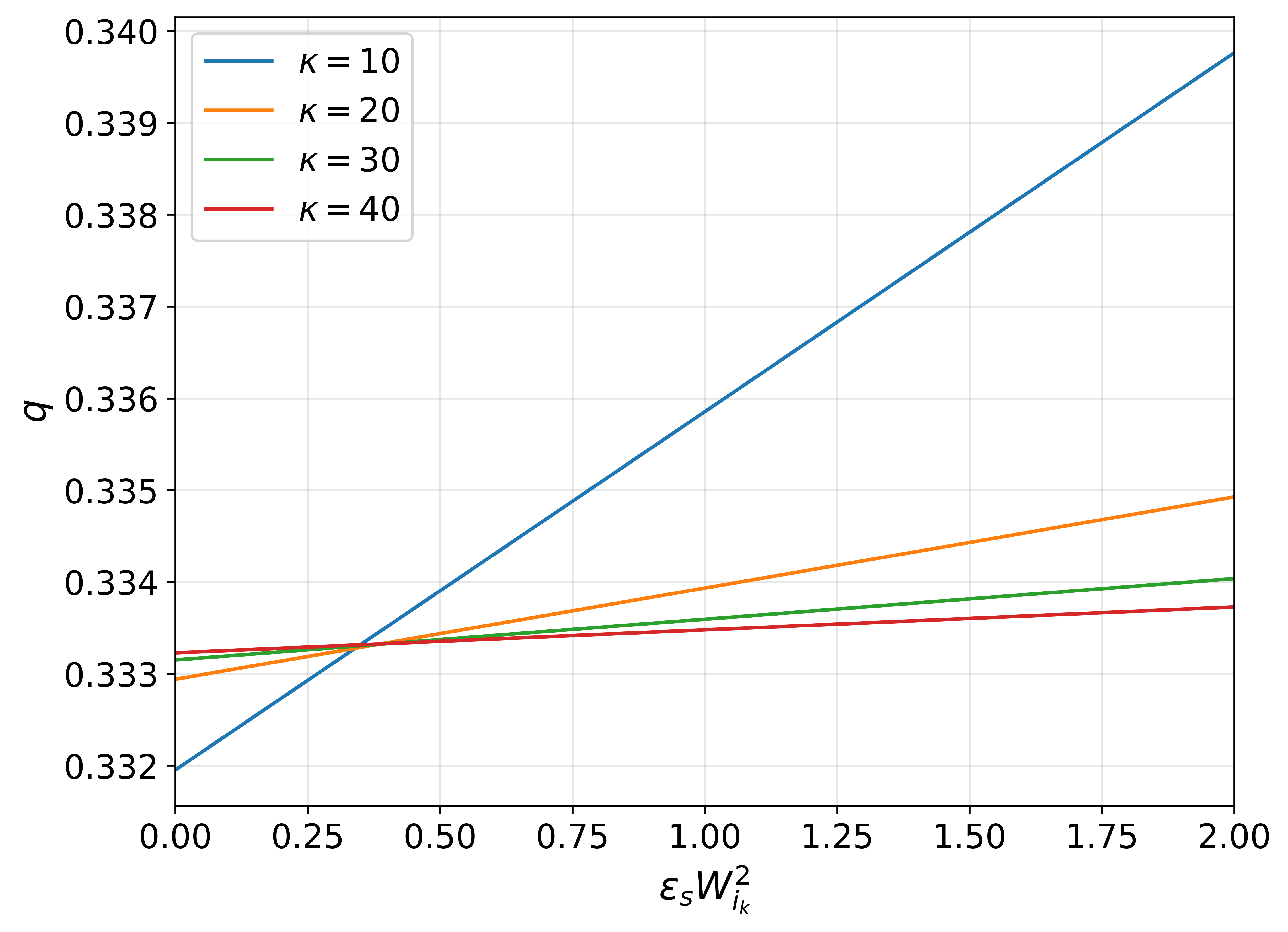}
    \caption{\(q\) vs \(\epsilon_s Wi_k^2\) plot from the surrogate outputs for \(\zeta = 4\), \(Du = 8\), and \(\kappa = 10, 20, 30, 40\)}
    \label{fig:q_vs_W}
\end{figure}

Figure~\ref{fig:q_vs_W} shows that the dimensionless volumetric flow rate \(q\) increases monotonically and almost linearly with \(\epsilon_s W_{i_k}^2\), with the steepest slope observed at \(\kappa = 10\).

This trend is a hallmark of shear-thinning rheology in the sPTT model. The exact slopes are close to \(2/(5\kappa^2)\) (e.g., \(4.0\times 10^{-3}\) for \(\kappa=10\)). At \(\epsilon_s Wi_k^2 = 0\) (Newtonian fluid), \(q\) lies below the Poiseuille value of \(1/3\) due to electroviscous retardation. As \(\epsilon_s Wi_k^2\) increases, the effective viscosity at high shear rates near the wall decreases, flattens the profile and increases the flow rate. The parameter combination \(c = 2\epsilon_s Wi_k^2/\kappa^2\) governs this effect, which is why the \(\kappa=10\) line is the steepest. The four lines cross near \(\epsilon_s Wi_k^2 \approx 0.35\) at \(q \approx 1/3\), which is the point where the shear-thinning gain exactly balances the electroviscous loss.

\subsubsection{Variation of Energy Conversion Efficiency with Viscoelastic Parameter} \label{sec:eta_vs_W}

\begin{figure}[htbp]
    \centering
    \includegraphics[width=0.9\columnwidth]{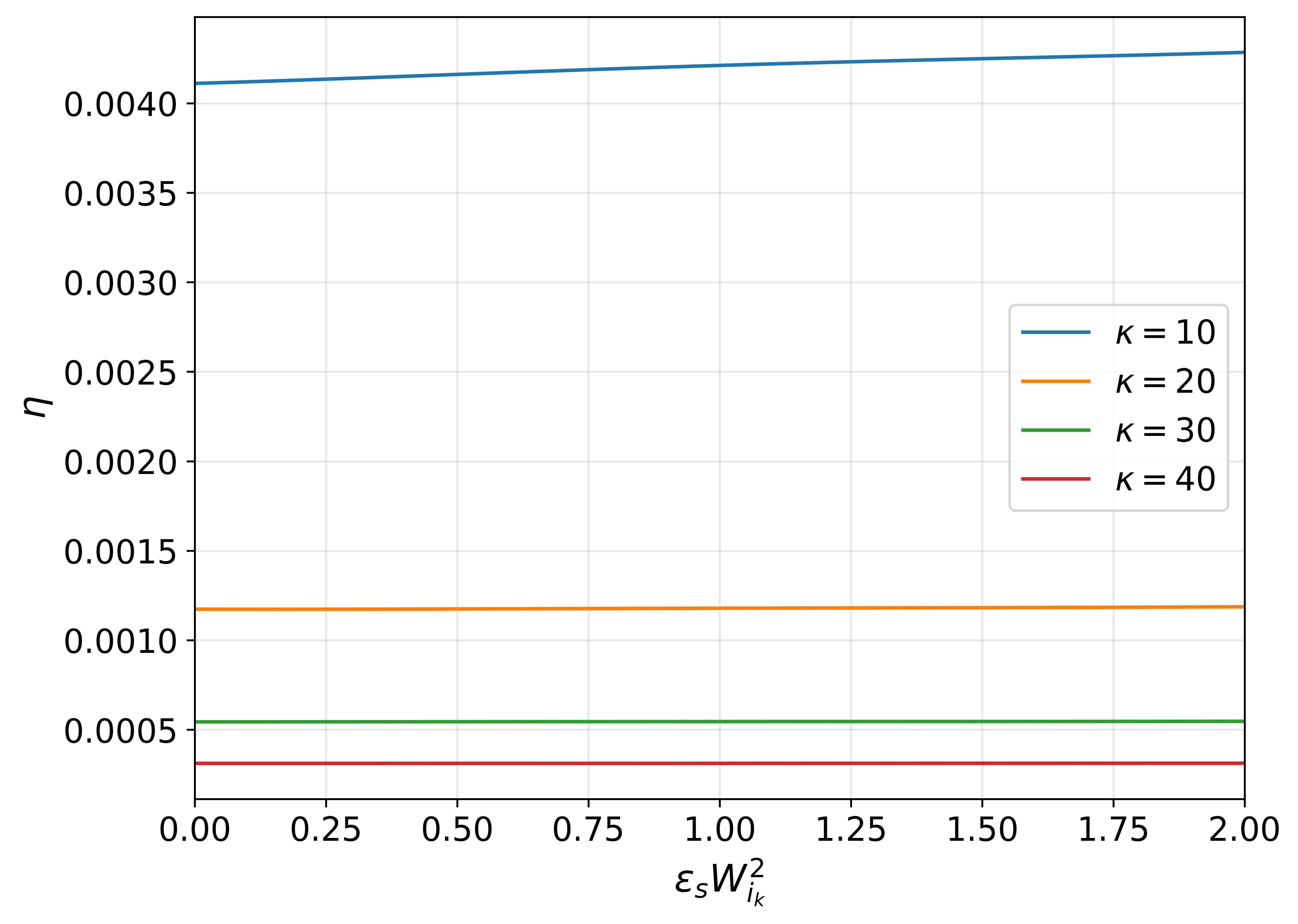}
    \caption{\(\eta\) vs \(\epsilon_s Wi_k^2\) plot from the surrogate outputs for \(\zeta = 4\), \(Du = 8\), and \(\kappa = 10, 20, 30, 40\)}
    \label{fig:eta_vs_W}
\end{figure}

Figure~\ref{fig:eta_vs_W} shows that the energy conversion efficiency \(\eta\) is nearly independent of the viscoelastic parameter \(\epsilon_s Wi_k^2\), remaining almost flat for all values of \(\kappa\) and changing by less than about \(5\%\).

As observed in Fig.~\ref{fig:eta_vs_W}, the near-independence of $\eta$ on $\epsilon_s W_{i_k}^2$ is consistent with the behavior of $E$ in Fig.~\ref{fig:E_vs_W}. The efficiency is given by \(\eta = (\kappa^2 \alpha/\zeta^2)(C+Du)E^2/q\). The parameter \(\epsilon_s W_{i_k}^2\) influences $\eta$ exclusively through its effect on \(E\) and \(q\). As \(E\) changes by only a few percent (as discussed in Fig.~\ref{fig:E_vs_W}) and \(q\) changes by less than $3\%$ for \(\kappa=10\), the efficiency remains nearly constant. Thus, enhanced viscoelasticity promotes a higher volumetric flow rate without a commensurate increase in conversion efficiency. The profile at \(\kappa=10\) originates from minor redistributions of the velocity field within the thicker \gls{edl}. This marginally alters the weighting of the streaming current integral and the conduction integral.

\subsubsection{Variation of Streaming Potential with Zeta Potential}

\begin{figure}[htbp]
    \centering
    \includegraphics[width=0.9\columnwidth]{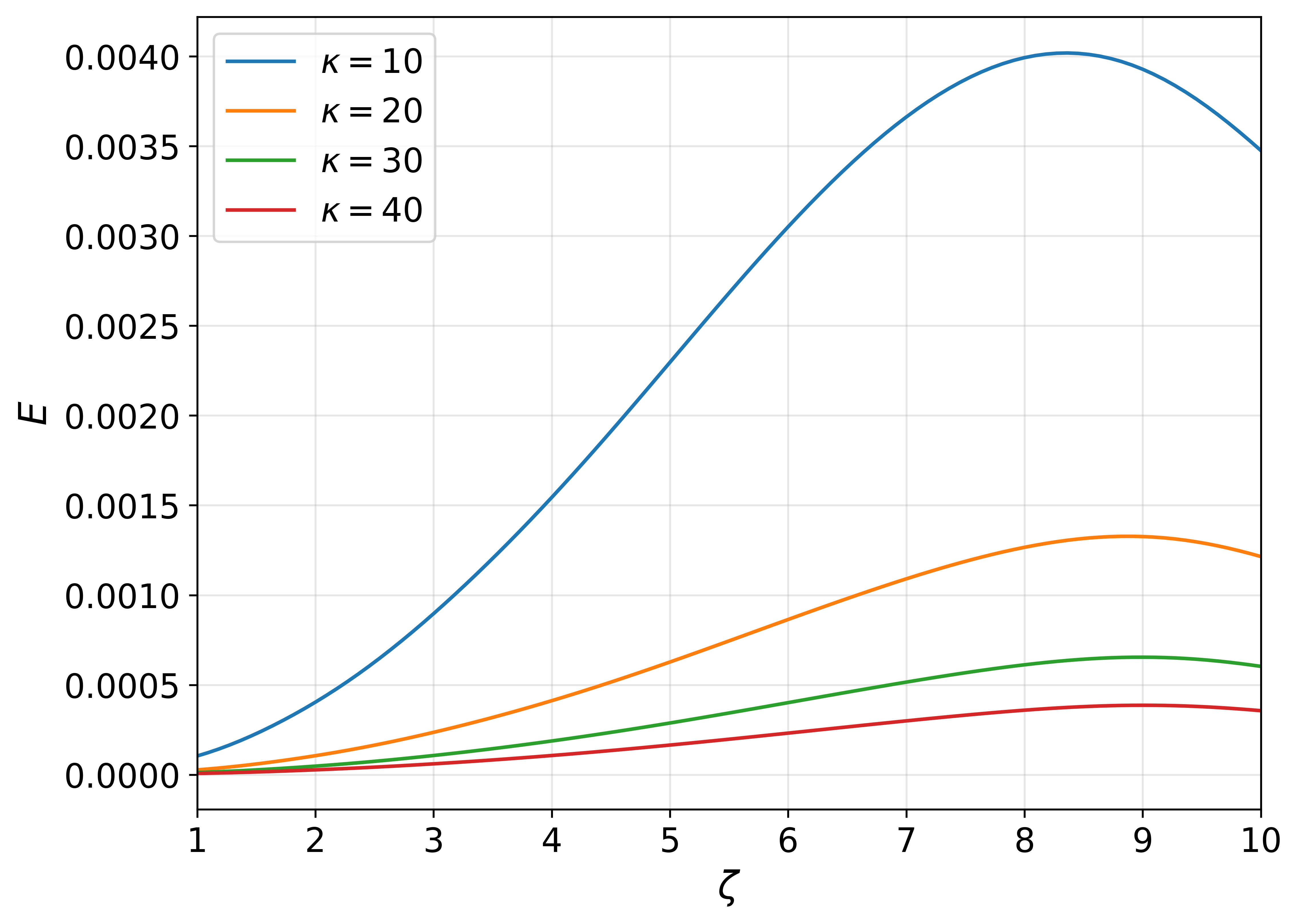}
    \caption{\(E\) vs \(\zeta\) plot from the surrogate outputs for \(\epsilon_s Wi_k^2 = 1.5\), \(Du = 8\), and \(\kappa = 10, 20, 30, 40\).}
    \label{fig:E_vs_zeta}
\end{figure}

As illustrated in Fig.~\ref{fig:E_vs_zeta}, streaming potential $E$ displays a bell-shaped dependence on the zeta potential $\zeta$, with the local of the maximum determined by both $Du$ and $\kappa$ (for e.g., peaking at $\zeta \approx 8.3$ for $\kappa=10$ and $Du=8$). Furthermore, the peak magnitude of $E$ decreases monotonically as $\kappa$ increases.

Initially, increasing \(\zeta\) increases the surface charge density and \(i_s\), causing \(E\) to grow roughly as \(\zeta^2\). However, the conduction integral \(C \approx 1 + (4/\kappa)\sinh^2(\zeta/4)\) grows exponentially as \(e^{\zeta/2}/\kappa\), short-circuiting the streaming current. The maximum is reached where the logarithmic growth rates of the numerator and denominator are equal (\(\frac{2}{\zeta} \approx \frac{C'(\zeta)}{C(\zeta)+Du}\)). The decline of \(E\) beyond this maximum is thus caused by the growth of the bulk conduction integral and not by counterion saturation (which the point-ion model does not contain). The entire curve scales down with increasing \(\kappa\) because thinner EDLs contain fewer mobile ions and lower velocities, causing both \(i_s\) and \(E\) to scale as \(1/\kappa^2\).

\subsubsection{Variation of Volumetric Flow Rate with Zeta Potential}

\begin{figure}[htbp]
    \centering
    \includegraphics[width=0.9\columnwidth]{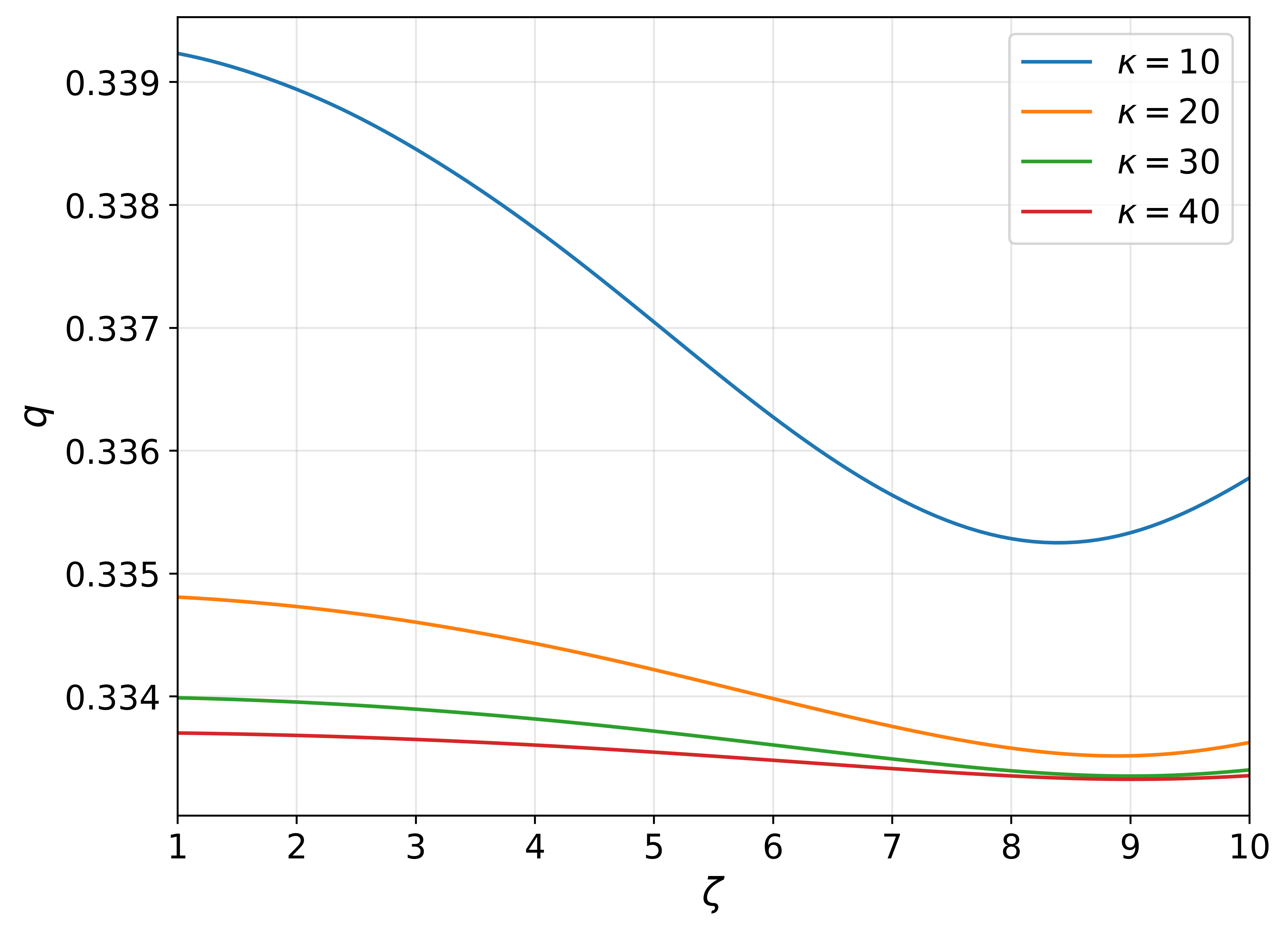}
    \caption{\(q\) vs \(\zeta\) plot from the surrogate outputs for \(\epsilon_s Wi_k^2 = 1.5\), \(Du = 8\), and \(\kappa = 10, 20, 30, 40\).}
    \label{fig:q_vs_zeta}
\end{figure}

Figure~\ref{fig:q_vs_zeta} shows that the flow rate $q$ shows a distinct minimum near $\zeta \approx 8.5$ (for $\epsilon_s W_{i_k}^2 = 1.5$ and $Du=8$) and subsequently recovers slightly at higher $\zeta$. The minimum is deepest for \(\kappa=10\) (about \(1.2\%\) below \(q_0\)).

This behavior is the mirror image of the streaming potential \(E\). The flow rate is reduced by the electrokinetic body force, which is proportional to \(E\). Since \(E\) peaks near \(\zeta \approx 8.5\), the retarding force is maximized there, causing \(q\) to dip to a minimum. At lower and higher values of \(\zeta\), \(E\) is smaller, which leads to less retardation and higher \(q\). The effect is most pronounced at \(\kappa=10\) because a thicker EDL allows the retarding body force to act over a larger cross-section. Importantly, electroviscous retardation acts at all \(\zeta\), and not only at large \(\zeta\). A higher zeta potential does not increase the flow rate.

\subsubsection{Variation of Energy Conversion Efficiency with Zeta Potential}

\begin{figure}[htbp]
    \centering
    \includegraphics[width=0.9\columnwidth]{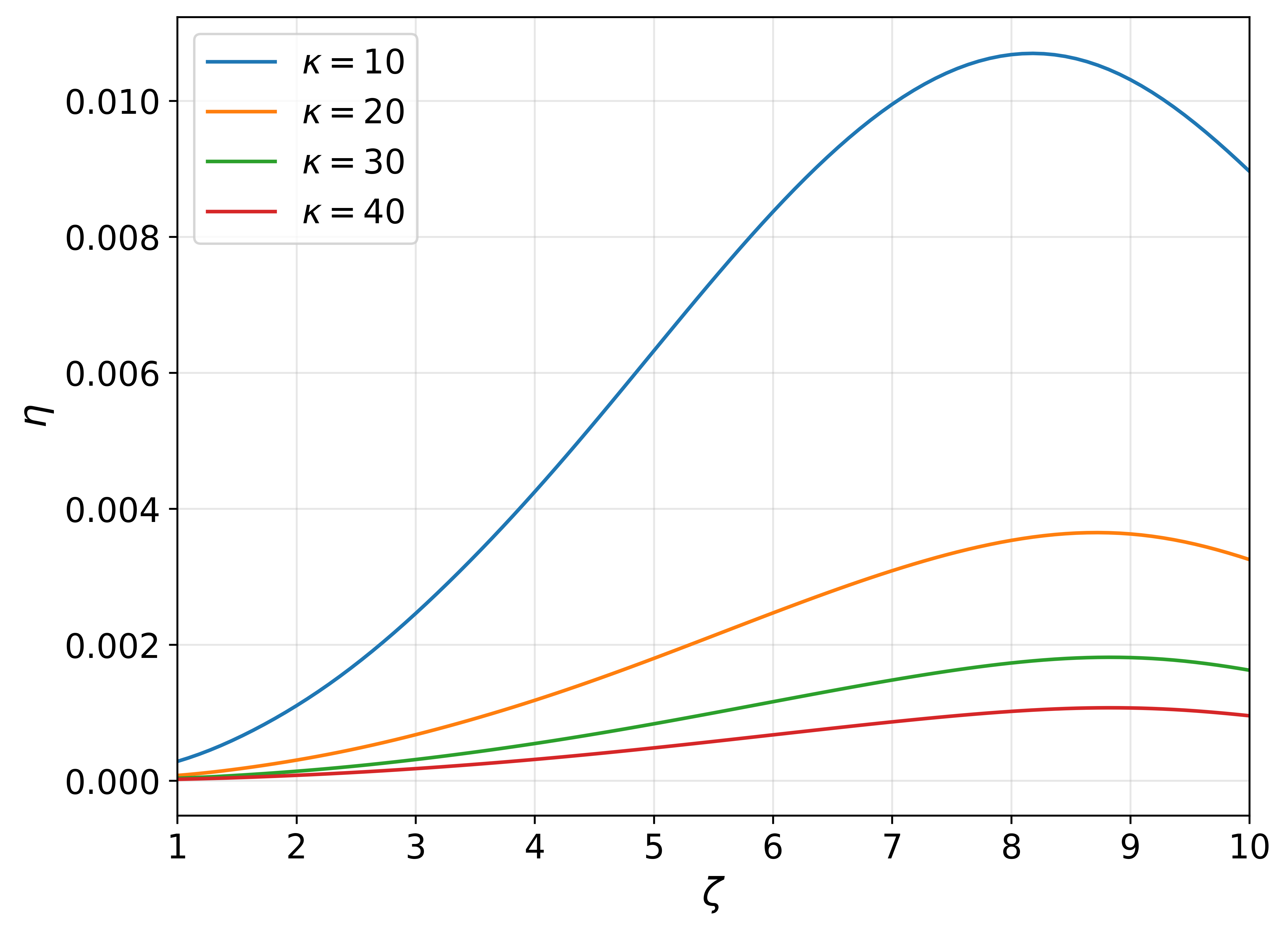}
    \caption{\(\eta\) vs \(\zeta\) plot from the surrogate outputs for \(\epsilon_s Wi_k^2 = 1.5\), \(Du = 8\), and \(\kappa = 10, 20, 30, 40\).}
    \label{fig:eta_vs_zeta}
\end{figure}

Figure~\ref{fig:eta_vs_zeta} presents the energy conversion efficiency $\eta$ as a function of the zeta potential $\zeta$. The conversion efficiency $\eta$ exhibits a distinct bell-shaped distribution with respect to the zeta potential $\zeta$, peaking near $\zeta \approx 8.5$ for all values of $\kappa$ (for $\epsilon_s W_{i_k}^2 = 1.5$ and $Du=8$). The magnitude of $\eta$ decreases monotonically as $\kappa$ increases.

As \(\zeta\) increases, the surface charge and counterion concentration in the EDL grow. This initially enhances the numerator of \(E\) (proportional to \(i_s \propto \zeta^2\)) and drives an increase in \(\eta\). However, beyond the optimal \(\zeta\), the denominator \(C \approx 1 + (4/\kappa)\sinh^2(\zeta/4)\) grows exponentially and acts as a conduction short-circuit. The efficiency peaks slightly below the maximum of \(E\) as \(\eta\) includes a factor of \(1/\zeta\) and \(i_s/\zeta\) decreases slowly with \(\zeta\). Regarding the \(\kappa\) dependence, higher \(\kappa\) denotes a thinner EDL. Both the streaming current \(i_s\) and the induced potential \(E\) scale as \(1/\kappa^2\), and therefore the efficiency falls by roughly an order of magnitude from \(\kappa=10\) to \(\kappa=40\).

\subsection{Pareto-Optimal Design Space}
\begin{figure}[h!]
    \centering
    \includegraphics[width=0.9\columnwidth]{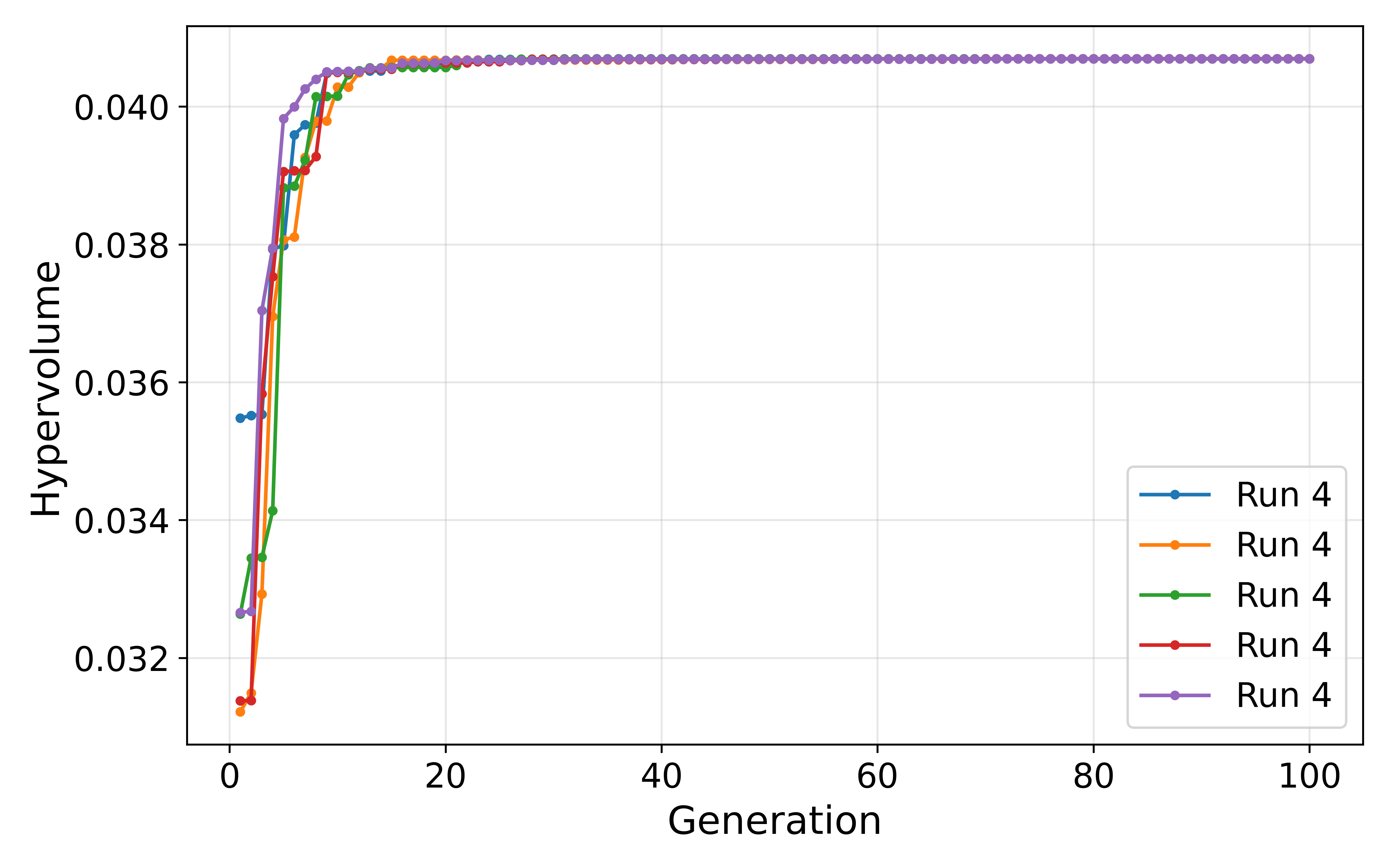}
    \caption{NSGA-II optimization history showing that the algorithm has converged within the chosen number (=100) of generations for all the five runs}
    \label{fig:nsga_history}
\end{figure}

The multi-objective search was performed with NSGA-II, using a population of 200 candidate designs evolved over 100 generations. This process was repeated 5 times with different random seeds for each run. To confirm that the reported front reflects a converged search rather than a result taken at an arbitrary stopping point, the hypervolume indicator of the non-dominated set was tracked across generations (Fig.~\ref{fig:nsga_history}). From Fig.~\ref{fig:nsga_history} it can be seen that it plateaued around generation 16 for all the five runs.

\begin{figure}[htbp]
    \centering
    \includegraphics[width=0.9\columnwidth]{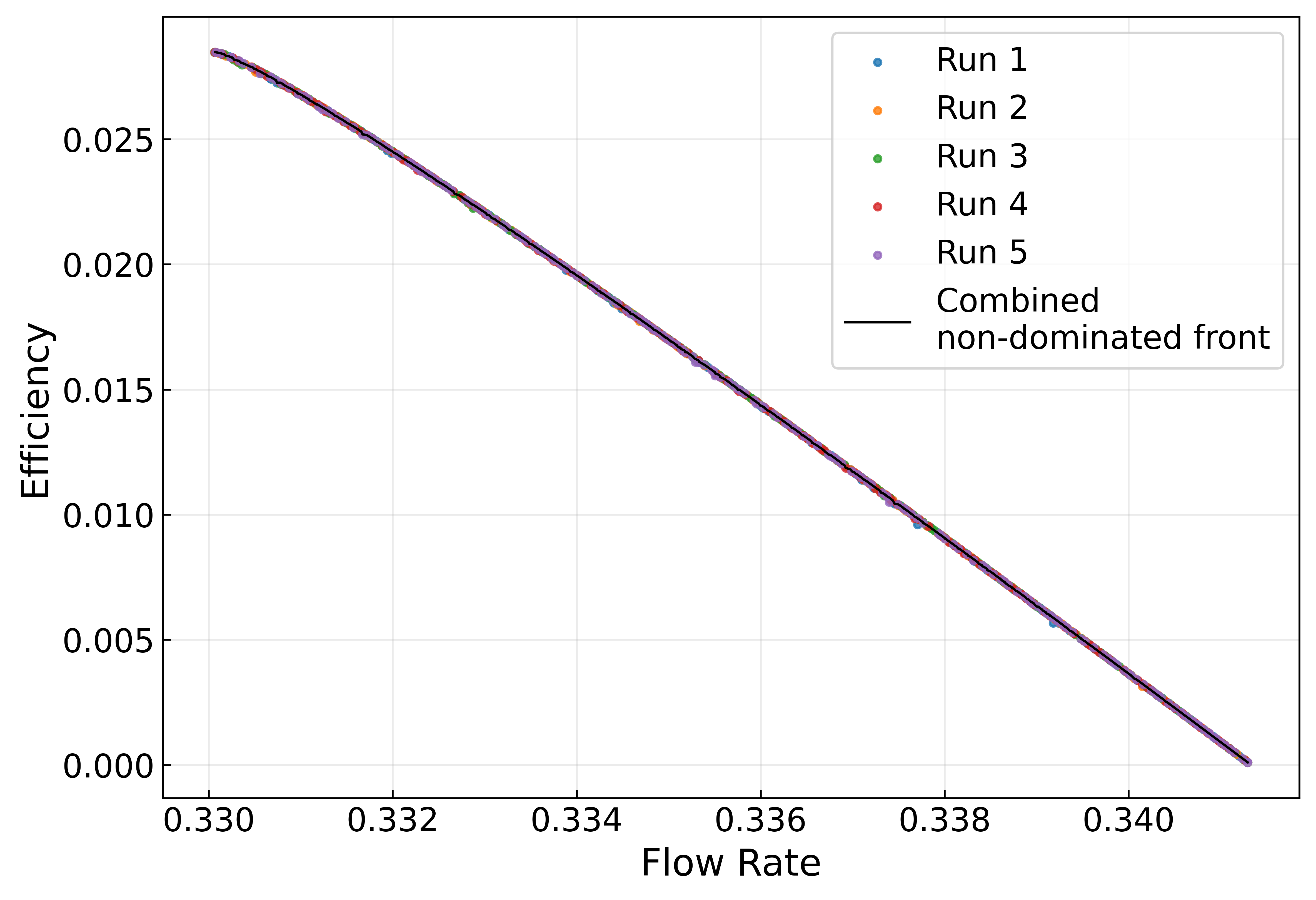}
    \caption{NSGA-II pareto front}
    \label{fig:nsga_pareto_tot}
\end{figure}

The NSGA-II search converges to a compact, well-defined Pareto front as shown in Fig.~\ref{fig:nsga_pareto_tot}. The Pareto fronts obtained from all the five runs overlap significantly indicating good consistency across different random initializations. The black line represents the overall Pareto front, obtained by performing non-dominated sorting on the combined set of solutions from all five runs.

To further assess the reliability of the surrogate model in the region relevant to multi-objective optimization, its predictive performance was evaluated over the non-dominated Pareto-optimal solutions. The outputs predicted by the \gls{nn}-surrogate were compared against those obtained from the original numerical solver. The performance metrics (mean L1 Error, \gls{rmse}, \gls{mxae}, \gls{mape}, and R2 score) for $E$, $q$, and $\eta$ are summarized in Table~\ref{tab:surrogate-err-pareto}. 

\begin{table*}[htbp]
\caption{Surrogate prediction errors for the non-dominated Pareto-optimal solutions, evaluated against the numerical solver.}
\label{tab:surrogate-err-pareto}
\begin{ruledtabular}
\begin{tabular}{ lccccc } 
 Output variable & Mean L1 & RMSE & MxAE & MAPE & R2 \\
\hline
    $E$ & 8.73452e-05 & 1.04870e-04 & 2.08907e-04 & 1.80255 & 0.99892 \\

    $q$ & 8.71046e-05 & 1.04579e-04 & 2.10124e-04 & 0.02602 & 0.99895 \\

    $\eta$ & 2.26269e-04 & 2.71336e-04 & 4.85159e-04 & 1.76176 & 0.99895 \\
\end{tabular}
\end{ruledtabular}
\end{table*}

From Table~\ref{tab:surrogate-err-pareto} The surrogate exhibits consistently high coefficients of determination ($R^2 > 0.998$) for all three quantities, indicating strong agreement with the numerical solver over the Pareto-optimal design space. The low Mean L1 and \gls{rmse} values further demonstrate that the surrogate maintains good predictive accuracy for the solutions identified by \gls{nsga2}. These results provide additional confidence that the identified Pareto front is not an artefact of surrogate prediction errors.

The Pareto front illustrates the trade-off between maximizing the dimensionless flow rate \(q\) and the energy conversion efficiency \(\eta\). The plot displays a nearly straight-line, competing relationship, i.e., designs that yield high flow rates correspond to lower efficiencies, and vice versa.
This behavior is a direct consequence of electroviscous retardation, where \(E\) controls both design objectives. The flow lost to retardation is \(q_0-q \approx E\), and the efficiency approximately follows \(\eta \approx A(q_0-q)\), with \(A\approx 2.6\) for \(\kappa=10\). The flow rate varies only slightly, whereas the efficiency decreases by more than two orders of magnitude across the Pareto front. Thus, a small variation in flow rate corresponds to a substantial change in efficiency. Since maximizing \(q\) requires minimizing the retarding body force (small \(E\)), while maximizing \(\eta\) requires a large \(E\), the optimizer maps a continuous Pareto front rather than converging to a single corner optimum.

\begin{figure*}[htbp]
    \centering

    \subfloat[$\kappa$\label{fig:pareto-kappa}]{%
        \includegraphics[width=0.48\textwidth]{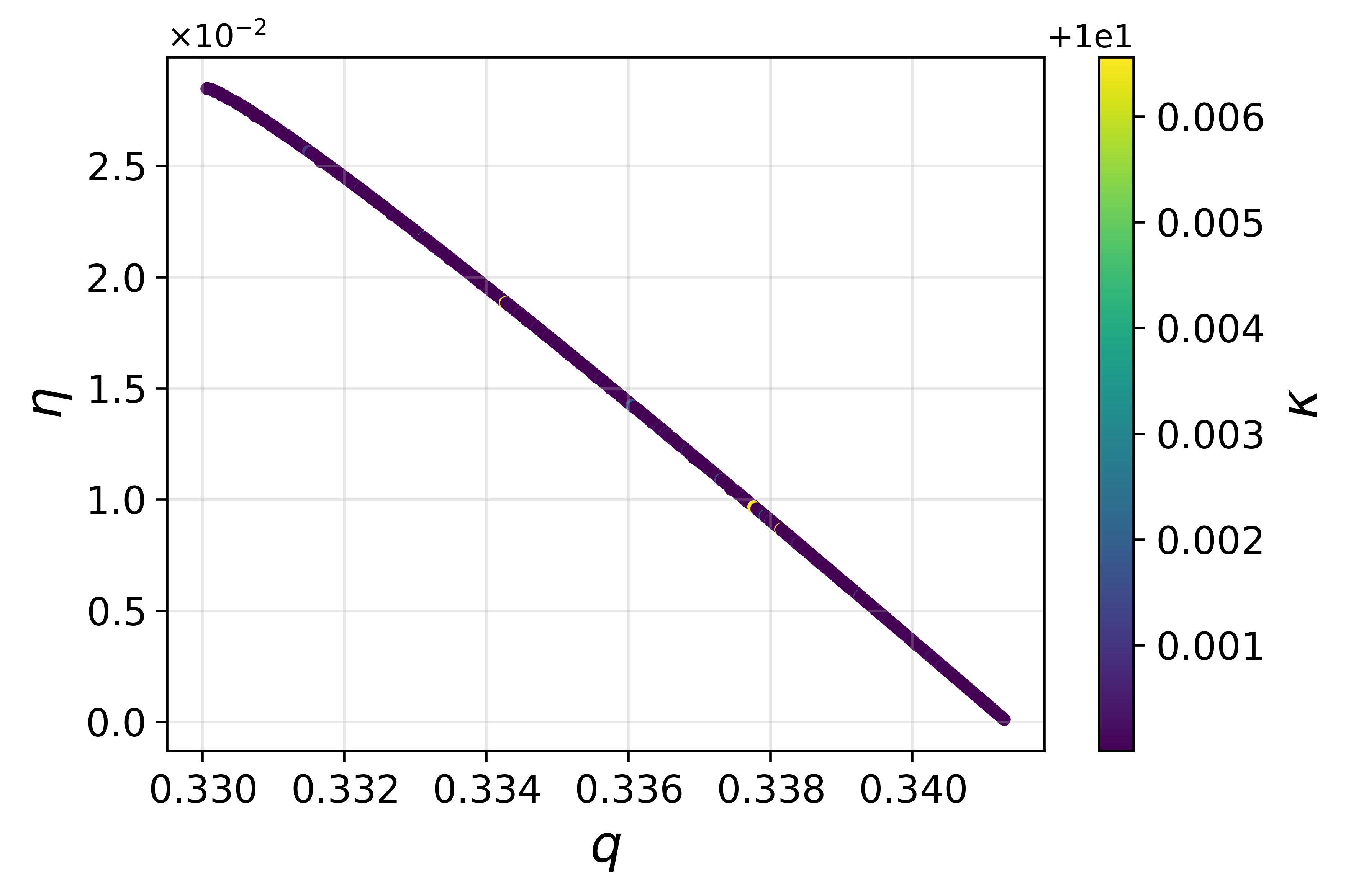}
    }
    \hfill
    \subfloat[$W$\label{fig:pareto-W}]{%
        \includegraphics[width=0.48\textwidth]{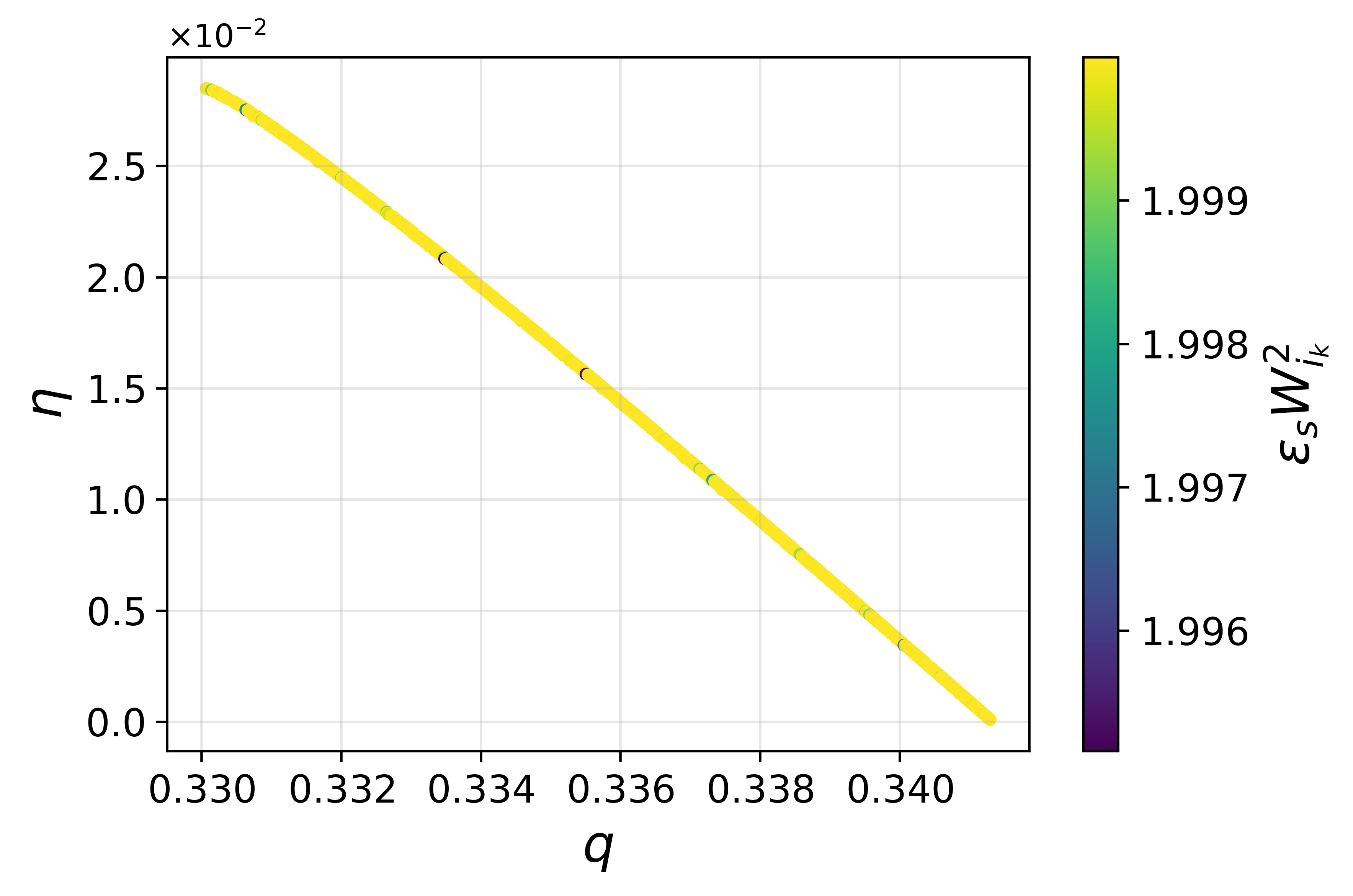}
    }

    \subfloat[$\zeta$\label{fig:pareto-zeta}]{%
        \includegraphics[width=0.48\textwidth]{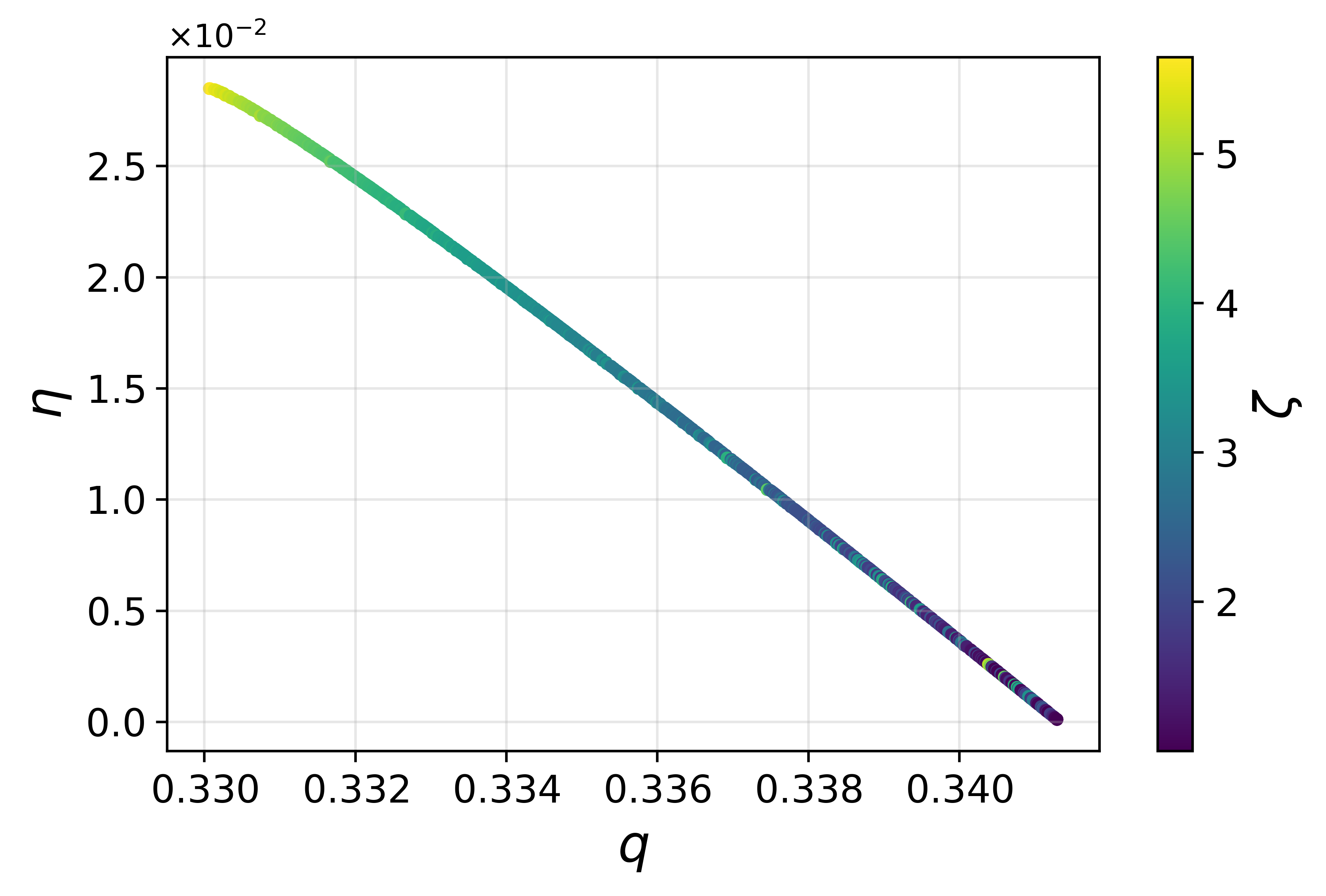}
    }
    \hfill
    \subfloat[$Du$\label{fig:pareto-Du}]{%
        \includegraphics[width=0.48\textwidth]{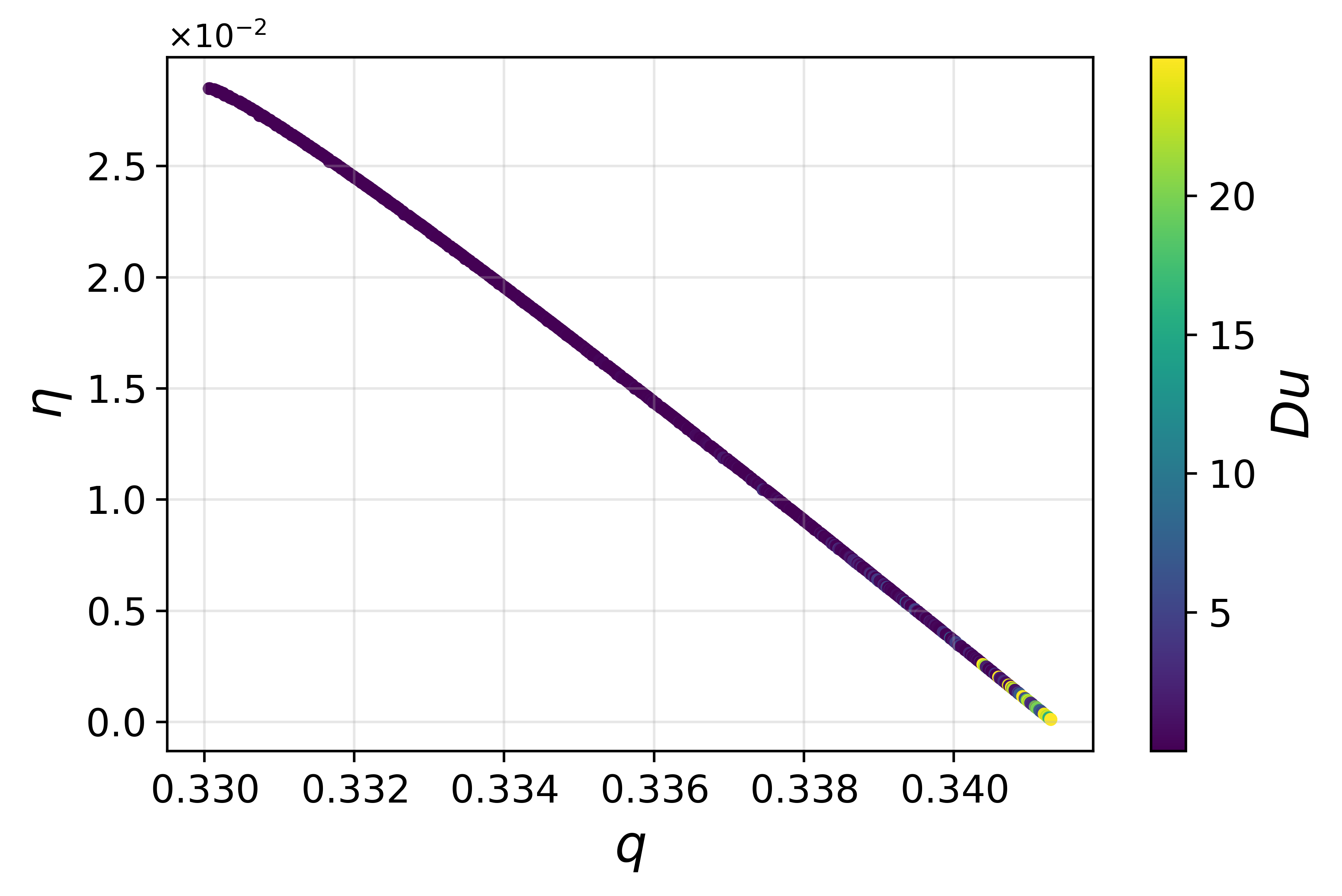}
    }

    \caption{NSGA-II Pareto fronts colored according to the values of the input parameters.}
    \label{fig:pareto-parameters}
\end{figure*}

The distributions of the input parameters among the non-dominated solutions are shown in Fig.~\ref{fig:pareto-parameters}. The results indicate that $\kappa$ remains close to its lower bound $\kappa \approx 10$, while $\epsilon_s W_{i_k}^{2}$ remains close to its upper bound ($\epsilon_s W_{i_k}^{2} \in [1.99516, 2]$). Consequently, these two parameters exhibit very little variation among the Pareto-optimal solutions. In contrast, $\zeta$ varies substantially along the Pareto front ($\zeta \in [0, 5.64725]$), decreasing as the volumetric flow rate increases and the efficiency decreases. The parameter $Du$ spans its entire range ($[0, 25]$) but remains close to its lower bound over most of the front and only increases towards the high-flow-rate, low-efficiency end.

The resulting Pareto front exhibits a clear trade-off between volumetric $q$ and $\eta$ --- solutions with higher $q$ are associated with lower $\eta$, and vice versa. The variation of $\zeta$ along the front indicates that it plays an important role in determining this trade-off. However, the variation of $Du$ near the high-flow-rate end suggests that the trade-off cannot be attributed exclusively to $\zeta$. Instead, both parameters contribute to the variation among the non-dominated designs, although their roles differ across the Pareto front.

These results extend the observations reported in Ref.~\cite{sarkarSpttStreamingPot}, which identified a favorable operating regime near $\zeta \approx 6$ and $\epsilon_s W_{i_k}^{2} \approx 2$ through parameter sweeps performed at fixed values of the other parameters. The present multiobjective optimization examines the combined influence of all four governing parameters and identifies their corresponding distributions across the Pareto-optimal region. In particular, it confirms the preference for high $\epsilon_s W_{i_k}^{2}$ and low $\kappa$, while revealing the roles of $\zeta$ and $Du$ in determining the balance between flow rate and efficiency. From a design perspective, the results indicate that $\kappa$ and $\epsilon_s W_{i_k}^{2}$ can be set close to their respective favorable bounds across the Pareto-optimal region. The remaining design choice primarily involves selecting $\zeta$, with $Du$ providing an additional degree of adjustment towards the high-flow-rate end.

\subsection{Global Sensitivity Analysis}

\begin{figure*}[htbp]
    \centering

    \subfloat[$q$\label{fig:sobol-q}]{%
        \includegraphics[width=0.4\textwidth]{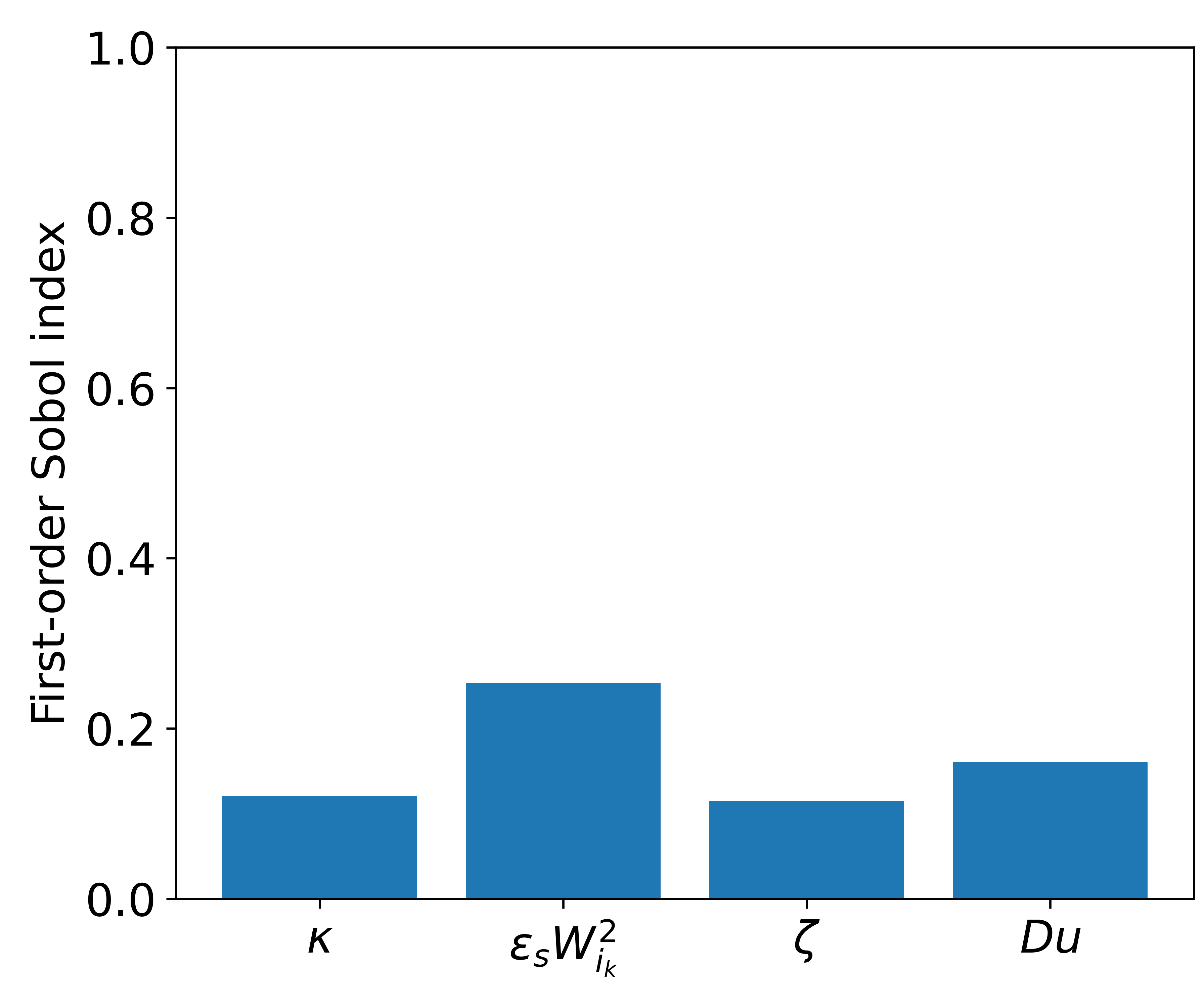}
    }
    \hfill
    \subfloat[$\eta$\label{fig:sobol-eta}]{%
        \includegraphics[width=0.4\textwidth]{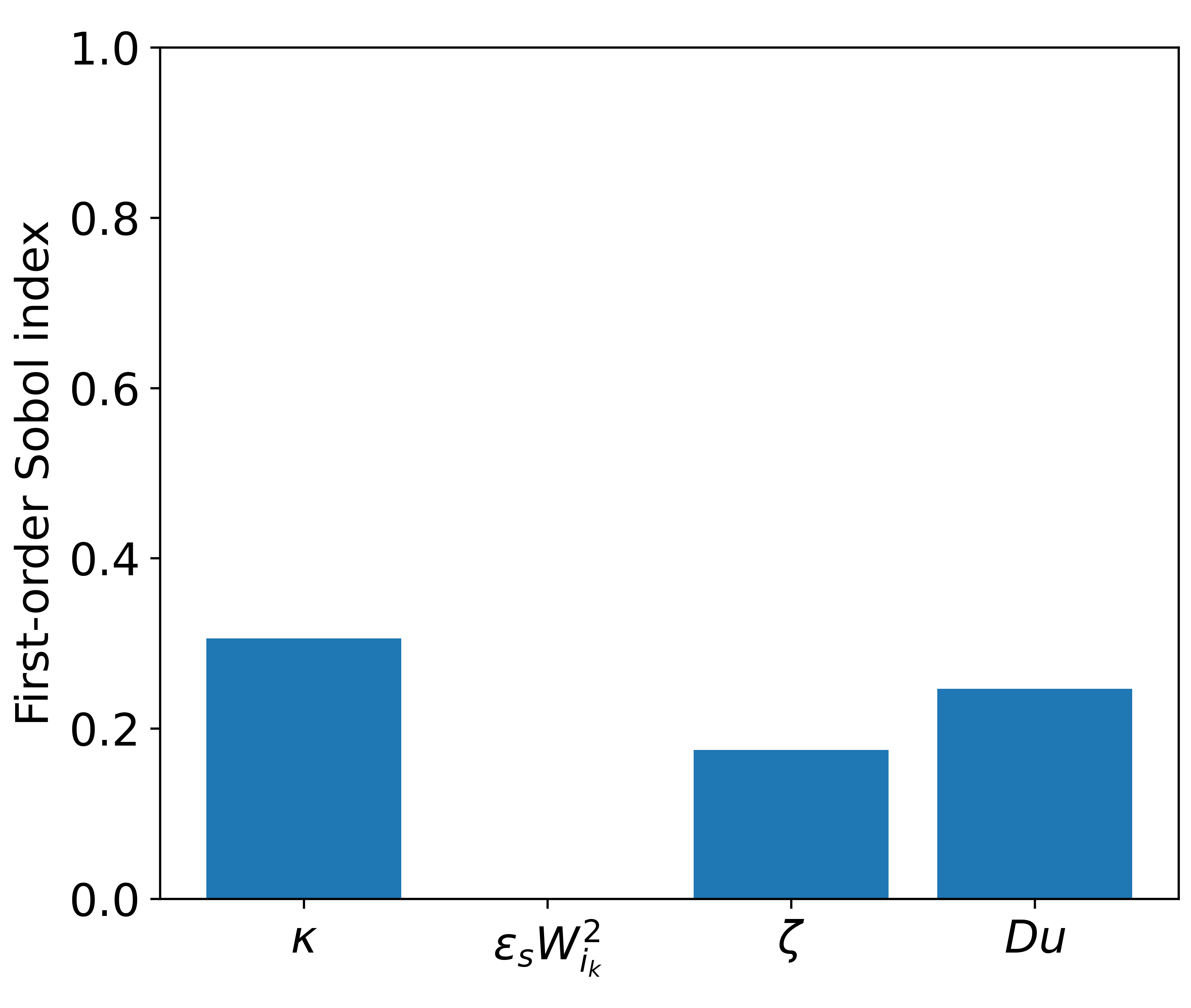}
    }

    \caption{First-order Sobol sensitivity analysis for flow rate ($q$) and efficiency ($\eta$).}
    \label{fig:sobol}
\end{figure*}

\begin{figure*}[htbp]
    \centering

    \subfloat[$q$\label{fig:sobol-q-total}]{%
        \includegraphics[width=0.4\textwidth]{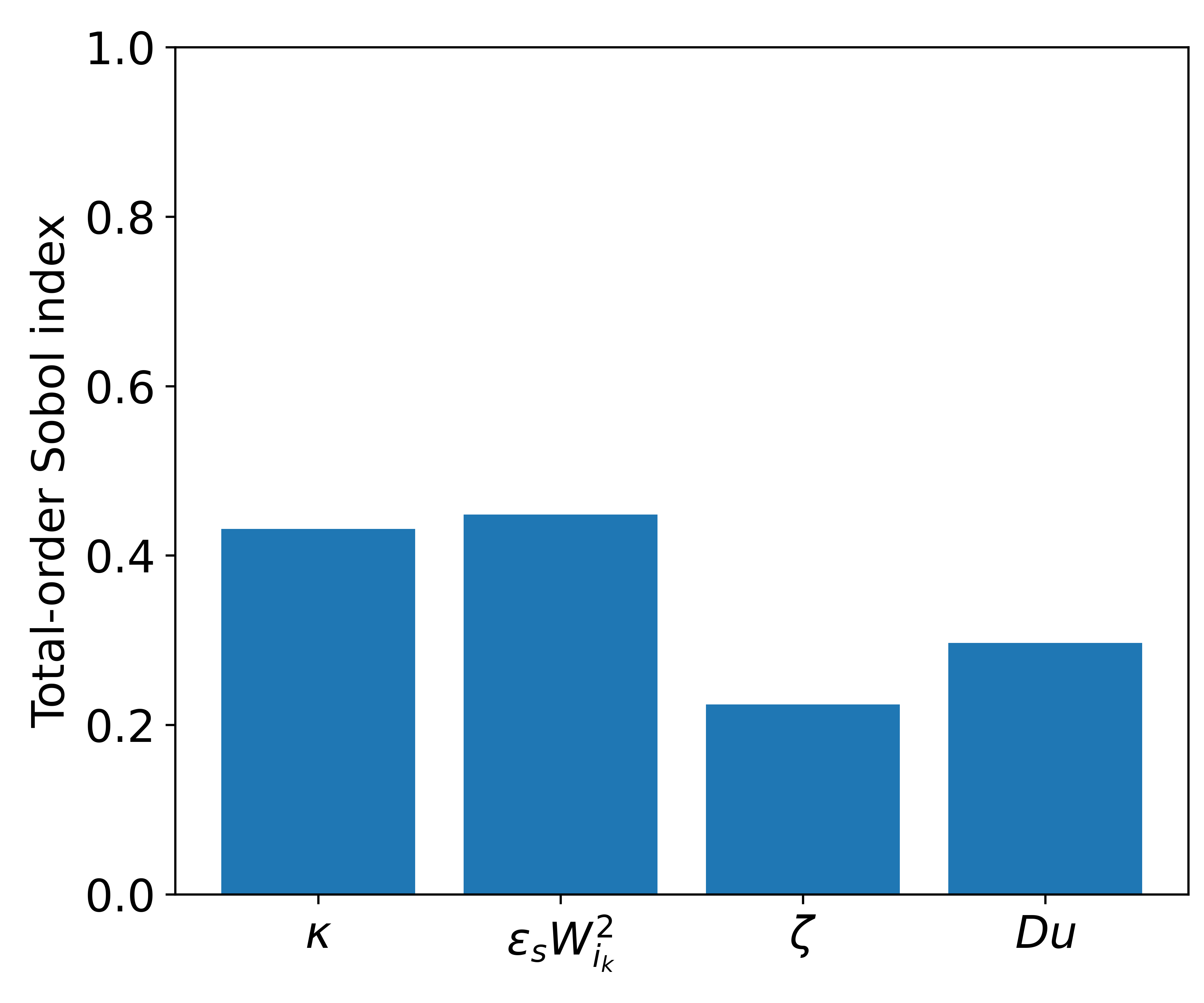}
    }
    \hfill
    \subfloat[$\eta$\label{fig:sobol-eta-tot}]{%
        \includegraphics[width=0.4\textwidth]{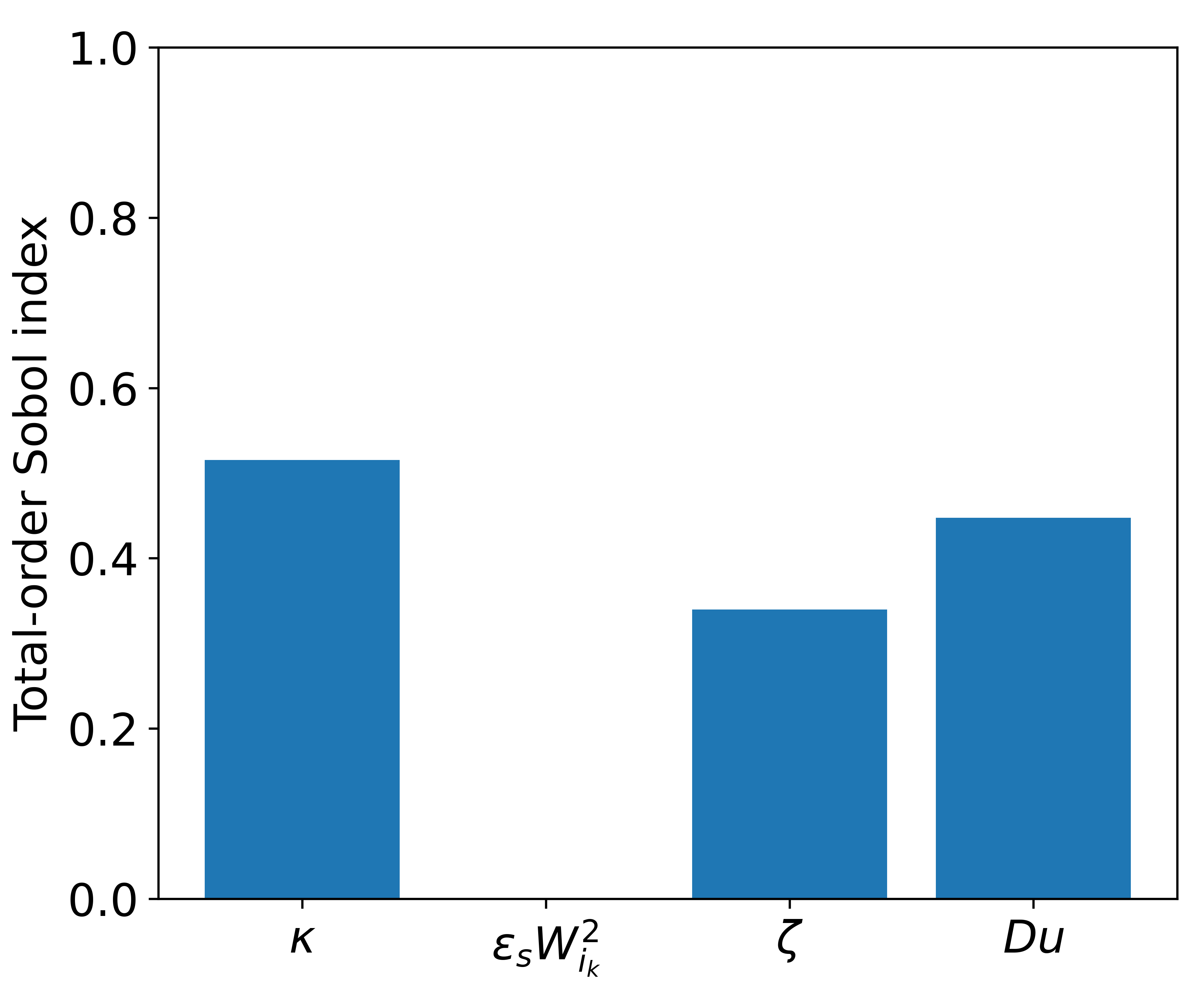}
    }

    \caption{Total-order Sobol sensitivity analysis for flow rate ($q$) and efficiency ($\eta$).}
    \label{fig:sobol-total}
\end{figure*}

To investigate the relative influence of the input parameters on the two optimization objectives, a global sensitivity analysis was performed.First-order and total-order Sobol sensitivity indices were computed for both objectives across the full search domain (Fig.~\ref{fig:sobol} and Fig.~\ref{fig:sobol-total}).

The first-order indices reveal distinct sensitivity patterns for the two objectives. For the volumetric flow rate ($q$), the viscoelastic parameter $\epsilon_s W_{i_k}^{2}$ exhibits the largest individual contribution to output variance ($S_1 \approx 0.25$), followed by $Du$ ($S_1 \approx 0.16$), $\kappa$ ($S_1 \approx 0.12$), and $\zeta$ ($S_1 \approx 0.11$). 
Physically, this sensitivity distribution for $q$ arises from the interaction between shear-thinning and electroviscous retardation. The flow rate can be approximated as $q \simeq q_0 - \gamma E$, where the zero-field shear-thinning flow rate is $q_0 = 1/3 + 2\epsilon_s Wi_k^2/(5\kappa^2)$, $\gamma \approx 0.9 - 1.0$ is the proportionality constant for electroviscous retardation, and $E$ is the streaming potential. The parameter $\epsilon_s W_{i_k}^2$ directly increases the baseline flow rate $q_0$, while $Du$ strongly suppresses the retarding streaming potential $E$ (as $E \propto 1/(C+Du)$). The parameter $\kappa$ has a dual effect --- it alters both $q_0$ and $E$ via $1/\kappa^2$ scaling.

In contrast, the efficiency ($\eta$) is most sensitive to $\kappa$ ($S_1 \approx 0.31$), followed by $Du$ ($S_1 \approx 0.25$) and $\zeta$ ($S_1 \approx 0.17$). The first-order index of $\epsilon_s W_{i_k}^{2}$ for efficiency is approximately zero, indicating a negligible individual contribution to the variance of this objective over the investigated domain. This is consistent with the findings of Section~\ref{sec:eta_vs_W}. 
The efficiency is rigorously defined as $\eta = \frac{\kappa^2 \alpha}{\zeta^2} \frac{(C+Du)E^2}{q}$, which depends on the \textit{streaming potential} $E$, the bulk conduction integral $C \approx 1 + \frac{4}{\kappa}\sinh^2(\zeta/4)$, and the flow rate $q$. The parameter $\kappa$ determines the \gls{edl} thickness. Smaller values of $\kappa$ increase both the streaming current $i_s$ and $E$, leading to higher efficiency. The Dukhin number $Du$ dictates the Stern layer conductivity --- as $Du$ increases, the term $(C+Du)$ provides a parallel conduction path that short-circuits $E$ and drastically reduces $\eta$. The negligible sensitivity to $\epsilon_s W_{i_k}^2$ arises because shear-thinning predominantly amplifies the core velocity while leaving the \gls{edl}-weighted streaming current nearly unchanged. Consequently, while $\epsilon_s W_{i_k}^2$ increases $q$, the ratio driving $\eta$ remains essentially invariant.

The total-order indices provide additional insight by accounting for all interactions involving each parameter. For $q$, the total-order indices of $\kappa$, $\epsilon_s W_{i_k}^{2}$, $\zeta$, and $Du$ are approximately $0.43$, $0.45$, $0.22$, and $0.30$, respectively. The differences between the first-order and total-order indices indicate that interactions contribute appreciably to the influence of these parameters, particularly $\kappa$. For efficiency, $\kappa$ has the largest total-order index ($S_T \approx 0.52$), followed by $Du$ ($S_T \approx 0.45$) and $\zeta$ ($S_T \approx 0.34$). The corresponding differences between $S_T$ and $S_1$ indicate substantial interaction contributions for all three parameters. In contrast, both indices for $\epsilon_s W_{i_k}^{2}$ remain approximately zero for efficiency, suggesting that its influence on this objective is negligible, both individually and through interactions, over the investigated domain.

\section{Conclusion}

In this work, a surrogate-assisted framework has been developed for the rapid design optimization of streaming-potential-mediated electrokinetic transport of \gls{sptt} fluids in a slit microchannel. The study addresses three critical gaps in the existing literature:

\begin{enumerate}
    \item \textbf{Surrogate modeling of implicit streaming potential}: Several data driver surrogates were trained to approximate the mapping \((\zeta, \kappa, \epsilon_s Wi_k^2, Du) \rightarrow E\), which aims to replace the computationally expensive iterative solution of the zero-net-current condition (Eq.~\eqref{eqn:electro-neutrality-nondim}). Among the surrogates, the neural network model was chosen as it achieved excellent accuracy, with L1 Loss of $1.61 \times 10^{-6}$ on the training set and $1.75 \times 10^{-6}$ on the testing set, which corresponds to percentage errors of 0.23\% and 0.24\%, respectively.
    
    \item \textbf{Global parametric exploration}: The trained surrogate enabled rapid evaluation of the entire four-dimensional design space, thereby revealing the governing parametric trends for the streaming potential \(E\), the volumetric flow rate \(q\), and the hydroelectric energy conversion efficiency \(\eta\). First-order Sobol sensitivity indices identified $\epsilon_s W_{i_k}^2$ as the dominant parameter for $q$ ($S_1 \approx 0.25$) and $\kappa$ for $\eta$ ($S_1 \approx 0.31$). The total order Sobol indices indicate substantial interaction contributions for all three parameters. However, both first-order and total-order indices for $\epsilon_s W_{i_k}^{2}$ remain approximately zero for $\eta$.
    
    \item \textbf{Multi-objective optimization}: The surrogate model was coupled with NSGA-II to perform a joint global search for operating conditions that simultaneously maximize \(q\) and \(\eta\). The optimization produced a well-defined Pareto front, with consistent results across five independent runs. The non-dominated solutions consistently favor the lower bound of \(\kappa\) (\(\kappa \approx 10\)) and the upper bound of \(\epsilon_s W_{i_k}^{2}\) (\(\epsilon_s W_{i_k}^{2} \approx 2\)). In contrast, \(\zeta\) varies substantially along the Pareto front (\(\zeta \in [0, 5.64725]\)), while \(Du\) spans its entire prescribed range (\([0,25]\)), remaining close to its lower bound over most of the front and increasing towards the high-flow-rate, low-efficiency region. These results identify \(\zeta\) and \(Du\) as important adjustable parameters governing the trade-off between flow rate and energy conversion efficiency.
\end{enumerate}

The physical insights derived from the multi-objective optimization are consistent with the fundamentals of electrokinetic theory, and are summarised as follows:
\begin{itemize}
    \item \textbf{Debye parameter ($\kappa$):} The optimizer converges to the lower bound ($\kappa \approx 10$). A smallar $\kappa$ corresponds to a thicker \gls{edl} and maximizes the number of mobile ions available for streaming current generation. This simultaneously enhances $E$ and $\eta$, and amplifies the shear-thinning gain in the flow rate.
    
    \item \textbf{Viscoelastic parameter ($\epsilon_s W_{i_k}^2$):}
    The optimzation favors the upperbound ($\epsilon_s W_{i_k}^2 \approx 2$). This maximizes shear-thinning and significantly enhances $q$. However, as the additional velocity from shear-thinning accumulates primarily in the channel core, rather than the thin \gls{edl}, the streaming potential and efficiency remain essentially invariant to this parameter. This is evidenced by its near-zero Sobol index for $\eta$.
    
    \item \textbf{Dukhin number ($Du$):} Efficiency is maximized when the stern layer conduction losses are eliminated ($Du \rightarrow 0$). As $Du$ increases, the highly conductive stern layer short-circuits the streaming potential, which monotonically decays $E$ and $\eta$. Conversely, this suppression of the retarding electrokinteic body force allows $q$ to increase towars its unretarded limit. 

    \item \textbf{Zeta potential ($\zeta$):} This serves as the primary adjustable parameter that traces the pareto front ($\zeta \in [0, 5.65]$). While, $E$ is bell-shaped with respect to $\zeta$, $q$ exhibits a corresponding minimum due to maximal electroviscous retardation at the peak of $E$. Consiquently, tuning $\zeta$ allows the system to transit smoothly between efficiency-optimal and flow-rate-optimal regimes.
\end{itemize}

Contrary to a single objective characterization, the analysis present a fundamental trade-off between $q$ and $\eta$ across the design space. The resulting pareto front is a nearly straight line where maximization of the flow rate necessitates a proportional sacrifice in energy conversion efficiency. This finding has practical implications for microfluidic device designs. Rather than a single optimal operating point, the designer must select a compromise along the pareto front dictated by the specific throughput and power requirements of the application.

The proposed surrogate methodology significantly accelerates this parametric exploration,  when compared with repeated numerical simulations. By reducing the computational cost of each objective-function evaluation from an iterative solution (with polylogarithmic quadrature) to a single forward pass through the neural network, rapid design space exploration becomes feasible. Hence, the findings provide actionable design guidelines for electrokinetic microfluidic devices:

\begin{enumerate}
    \item Operating at the lowest possible $\kappa$ (within the thin-\gls{edl}fabrication constraints) to maximize \gls{edl} thickness and simultaneously boost efficiency and flow rate;
    \item Utilizing fluids with the highest possible $\epsilon_s W_{i_k}^2$ (stronger viscoelasticity) to exploit shear-thinning enhancement for throughput without degrading $\eta$;
    \item Minimizing Stern layer conductivity ($Du \rightarrow 0$) through surface treatments or channel material selection to maximize the induced $E$ and $\eta$;
    \item Designing the system with a tuned $\zeta$ to balance the tradeoff between throughput and efficiency.
\end{enumerate}

\section{Limitations and Scope of Validity} \label{sec:limitations}
The findings of this study are subject to the following constraints and should be interpreted accordingly:

\begin{enumerate}
    
    \item \textbf{Bounded design space:} All results are confined to the parameter ranges established in Ref.~\cite{sarkarSpttStreamingPot} (\(\zeta \in [1,10]\), \(\kappa \in [10,40]\), \(\epsilon_s W_{ik}^2 \in [0,2]\), \(Du \in [0,25]\)), which were deliberately retained to maintain consistency with the underlying semi-analytical model. As two of the Pareto-optimal coordinates (\(\kappa \approx 10\) and \(\epsilon_s W_{ik}^2 \approx 2\)) lie on the boundaries of this domain, they represent the best designs within the studied range rather than unconstrained global optima, and should not be extrapolated beyond these limits. In particular, the lower bound on \(\kappa\) is tied to the model's assumptions --- Eq.~eqref{eqn:edl-pot-dimensionless} represents an isolated-wall solution that satisfies the centerline symmetry condition only up to terms of order \(e^{-\kappa}\). While the centreline potential is below \(2 \times 10^{-4}\) at \(\kappa = 10\) (validating the thin-\gls{edl} approximation), this assumption, alongside the non-overlapping \gls{edl} premise, becomes inadequate as \(\kappa\) approaches order unity. Consequently, the guideline to operate at low \(\kappa\) is strictly applicable only within the thin-EDL regime.

    \item \textbf{Point-ion Poisson-Boltzmann model:} The ionic distribution assumes a Boltzmann law without steric (finite-ion-size) effects. At \(\zeta = 10\), the counterion concentration at the wall is of order \(e^{\zeta} \approx 2 \times 10^4\) times the bulk value, which exceeds physical packing limits for millimolar electrolytes. Therefore, results in the upper region of the \(\zeta\) range---including the location and height of the maximum of \(E(\zeta)\)--- are indicative rather than strictly quantitative. Within this point-ion framework, the decline of \(E\) at large \(\zeta\) arises from the growth of the bulk conduction integral and not from counterion saturation (a phenomenon that a point-ion description cannot represent). Furthermore, the conductivity parameter was held fixed at \(\alpha = 10\) throughout the analysis.

    \item \textbf{Flow model:} The analysis assumes steady, fully developed, unidirectional, isothermal flow of a single-mode sPTT fluid with constant properties. It assumes no wall slip, negligible solvent viscosity (\(\eta_{\text{eff}} = 0\)), and a slit geometry with \(W \ll L\). Entrance effects, Joule heating, polymer-depleted layers, and the multi-mode relaxation of real biofluids are not captured. Thus, the results serve as model-based design guidance rather than quantitative predictions for a specific fluid, and no experimental validation is attempted.

    \item \textbf{Independence of dimensionless groups:} The four dimensionless groups are treated as independent decision variables, whereas in a physical device they are coupled through the half-height \(a\), the ionic strength \(n_0\), and the polymer relaxation time. For instance, \(W_{ik} = \lambda \kappa U_{ref} / a\), so changing \(\kappa\) at a fixed \(\epsilon_s W_{ik}^2\) implies a concurrent change in \(\lambda\) or \(U_{ref}\). Consequently, the Pareto set constitutes a dimensionless design map; an optimization performed in dimensional variables with these physical couplings imposed could yield a different set of optimal designs. For context, for a 1:1 electrolyte at \(25^{\circ}\text{C}\), the Debye length is \(\lambda_D \approx 0.304 / \sqrt{C}\) nm (with \(C\) in mol/L). Thus, \(\kappa \in [10, 40]\) corresponds to half-heights of roughly 0.1--0.4 \(\mu\)m at 1 mM, 0.3--1.2 \(\mu\)m at 0.1 mM, and only 8--31 nm at physiological ionic strength (150 mM). Similarly, the optimal \(\zeta \approx 6\) (approximately \(150\text{ mV}\)) lies toward the upper limit of zeta potentials typically sustained by microchannel walls.

    \item \textbf{Objective functions:} Because the dimensionless flow rate \(q\) is normalized by the Newtonian Poiseuille scale, it varies by only a few percent across the design space (\(q_0 - q \approx E\)). Therefore, the trade-off is characterized by a nearly constant flow rate competing against an efficiency that varies over orders of magnitude. Moreover, following Ref.~\cite{sarkarSpttStreamingPot}, \(\eta\) is evaluated under the zero-net-current condition. It serves as a figure of merit for hydroelectric conversion rather than the efficiency of power delivered to an external load; load matching is not modeled.

    \item \textbf{Surrogate scope:} The Feed-Forward Neural Network (FFNN) was trained exclusively on data generated by this model and is valid only within the sampled design box; no extrapolation is intended. Its accuracy is lower near the boundaries of the design space---where the Pareto-optimal designs reside---compared to the random test set (yielding a MAPE of approximately 1.8\% for \(E\) at the Pareto points versus 0.24\% on the test set), because Latin Hypercube sampling is sparsest in these regions. To ensure reliability, the Pareto points were re-evaluated using the numerical solver (refer to Table~\ref{tab:surrogate-err-pareto}). The Sobol indices and Pareto sets are computed via the surrogate and inherently inherit its prediction error. The surrogate replaces only the implicit evaluation of \(E\), whereas \(q\) and \(\eta\) still require quadrature of closed-form expressions; the reported computational speed-up refers specifically to the iterative solver used here. A solver that exploits the polynomial dependence of \(u\) on \(E\) would narrow this gap. Finally, the one-time cost of data generation and training is of the same order as a single direct run of the optimizer, meaning the surrogate yields substantial returns primarily when reused for repeated searches, parametric sweeps, and sensitivity analyses.

    \item \textbf{Hyperparameter choice:} The network architecture and hyperparameters were not systematically optimized. The accuracy of the surrogate was verified directly against the numerical solver on an independent test set (Table~\ref{tab:surrogate-res}) and at the Pareto-optimal designs (Table~\ref{tab:surrogate-err-pareto}), and further tuning was not considered necessary for the purposes of this study. So, the accuracy reported here does not necessarily represent the best attainable accuracy.
\end{enumerate}

\section{Future Work}

To address the limitations outlined in Section~\ref{sec:limitations}, future investigations should focus on several key theoretical and practical extensions. A natural progression would involve extending the theoretical framework to smaller \(\kappa\) using a finite-slit, overlapping-EDL Poisson-Boltzmann solution, and to larger \(\zeta\) using a modified Poisson-Boltzmann model that accounts for finite ion size. This would establish whether the observed parametric trends and the location of the optima persist beyond the currently established bounds. 

Furthermore, performing the multi-objective optimization in dimensional variables (such as the bulk ionic concentration \(n_0\), imposed pressure drop, wall chemistry, and polymer concentration) while explicitly enforcing the physical couplings discussed previously would translate the dimensionless Pareto set into practically realizable device designs. The underlying flow model could also be enriched by incorporating finite solvent viscosity, thermal effects (including Joule heating and Soret effects), wall slip, polymer-depleted layers, and alternative viscoelastic constitutive equations (e.g., Giesekus or FENE-P). Concurrently, the energy conversion efficiency could be evaluated under finite external loads rather than the zero-net-current condition, and the optimization could be expanded to include additional design objectives, such as the minimization of polymeric stresses.

On the computational and data-driven front, implementing ensemble or dropout-based uncertainty quantification, coupled with active learning to refine the training database near the optimal Pareto front, would systematically reduce surrogate prediction errors in critical regions. A systematic hyperparameter and architecture study (network depth and width, activation function, and learning-rate schedule, for example by Bayesian optimization) could further reduce the surrogate error near the boundaries of the design space. Additionally, benchmarking the current surrogate approach against exact reduced-order solvers or hybrid \glspl{pinn} for field reconstruction could provide further computational advantages. Finally, to validate the predicted electrokinetic and rheological trends, future efforts must include experimental measurements of the streaming potential and volumetric flow rate for viscoelastic fluids in microchannel geometries.

\section*{Supplementary Information}
\subsection*{Authors' contributions}
\noindent
\textbf{A.B.:} Conceptualization, Data Curation, Formal Analysis, Investigation, Methodology, Software, Validation, Visualization, Writing -- Original Draft, Writing -- Review \& Editing. \\
\textbf{S.B.:} Conceptualization, Formal Analysis, Investigation, Methodology, Project Administration, Supervision, Validation, Visualization, Writing -- Original Draft, Writing -- Review \& Editing. \\
\textbf{S.S.:} Conceptualization, Supervision.

\subsection*{Funding}
Not applicable

\subsection*{Data availability}
Research data have been generated in this study and made publicly in the following repository: \url{https://github.com/Ankan-Basu/Streaming-Potential-Optimisation}

\subsection*{Declarations}

\subsubsection*{Ethics statement}
Not applicable.

\subsubsection*{Competing interests}
The authors declare that they have no competing interests

% \nocite{*}
\bibliography{refs}% Produces the bibliography via BibTeX.

\end{document}